\documentclass[format=acmsmall, review=false, screen=true, authorversion]{acmart}

\usepackage{booktabs} %

\usepackage{algorithm}
\usepackage{algpseudocode}

\usepackage{tabularx}

\usepackage{xspace}
\newcommand\workflow{workflow}
\newcommand\Workflow{Workflow}
\newcommand\module{module}

\newcommand\extract{extract}
\newcommand\Extract{Extract}
\newcommand{\QAforML}{QA4ML\xspace}

\setcopyright{acmlicensed}
\copyrightyear{2023}
\acmYear{2023}
\acmPrice{15.00}
\acmDOI{10.1145/3594552}

\acmJournal{TIIS}
\acmVolume{}
\acmNumber{}
\acmArticle{}
\acmMonth{5}

\begin{document}
\title[Simulation-Based Optimization of QA4ML Interfaces]{Simulation-Based Optimization of User Interfaces for Quality-Assuring Machine Learning Model Predictions}

\author{Yu Zhang}
\orcid{0000-0002-9035-0463}
\affiliation{
  \institution{University of Oxford}
  \city{Oxford}
  \postcode{OX1 3QD}
  \country{The United Kingdom}
}
\email{yu.zhang@cs.ox.ac.uk}

\author{Martijn Tennekes}
\orcid{0000-0002-6506-9522}
\affiliation{
  \institution{Statistics Netherlands}
  \city{Heerlen}
  \postcode{6401 CZ}
  \country{The Netherlands}
}
\email{m.tennekes@cbs.nl}

\author{Tim de Jong}
\orcid{0000-0001-7845-9602}
\affiliation{
  \institution{Statistics Netherlands}
  \city{Heerlen}
  \postcode{6401 CZ}
  \country{The Netherlands}
}
\email{tja.dejong@cbs.nl}

\author{Lyana Curier}
\orcid{0000-0002-9108-3539}
\affiliation{
  \institution{Open University of the Netherlands}
  \city{Heerlen}
  \postcode{6401 DL}
  \country{The Netherlands}
}
\email{lyana.curier@ou.nl}

\author{Bob Coecke}
\orcid{0000-0002-5310-8723}
\affiliation{
  \institution{University of Oxford}
  \city{Oxford}
  \postcode{OX1 3QD}
  \country{The United Kingdom}
}
\email{bob.coecke@cs.ox.ac.uk}

\author{Min Chen}
\orcid{0000-0001-5320-5729}
\affiliation{
  \institution{University of Oxford}
  \city{Oxford}
  \postcode{OX1 3QD}
  \country{The United Kingdom}
}
\email{min.chen@eng.ox.ac.uk}

\renewcommand\shortauthors{Y. Zhang et al.}

\begin{abstract}
  
Quality-sensitive applications of machine learning (ML) require quality assurance (QA) by humans before the predictions of an ML model can be deployed.
QA for ML (\QAforML) interfaces require users to view a large amount of data and perform many interactions to correct errors made by the ML model.
An optimized user interface (UI) can significantly reduce interaction costs.
While UI optimization can be informed by user studies evaluating design options, this approach is not scalable because there are typically numerous small variations that can affect the efficiency of a \QAforML interface.
Hence, we propose using simulation to evaluate and aid the optimization of \QAforML interfaces.
In particular, we focus on simulating the combined effects of human intelligence in initiating appropriate interaction commands and machine intelligence in providing algorithmic assistance for accelerating \QAforML processes.
As \QAforML is usually labor-intensive, we use the simulated task completion time as the metric for UI optimization under different interface and algorithm setups.
We demonstrate the usage of this UI design method in several \QAforML applications.

\end{abstract}

\ccsdesc[500]{Human-centered computing~HCI design and evaluation methods}
\ccsdesc[300]{Information systems~Multimedia information systems}

\keywords{model-based evaluation, quality assurance, interactive machine learning, data labeling, classification}

\ifx\hidemain\undefined
  \received{25 October 2022}
  \received[revised]{26 February 2023}
  \received[accepted]{02 April 2023}

  \maketitle

  \section{Introduction}

\textbf{Machine learning (ML)} models rarely achieve 100\% prediction accuracy in real-world applications.
In some applications, such as search engines and product recommendations, the end-users can ignore prediction errors with little cost, and thus the errors are tolerable.
However, prediction errors cannot be ignored in many other applications, such as medical diagnosis, fraud detection, and content moderation.
Such applications are characterized as \emph{quality-sensitive}, where rigorous \textbf{quality assurance (QA)} of model predictions is necessary.

During QA processes, human needs to review machine predictions and correct erroneous predictions in a \textbf{user interface (UI)}.
This work focuses on interfaces for performing such repetitive tasks of quality-assuring ML model predictions.
We refer to these interfaces as \textbf{QA for ML (\QAforML)} interfaces.

In the software development lifecycle of an interactive system, it is important to explore and evaluate many different design options to select a relatively optimized design to be implemented, tested, and deployed.
The design optimization process can also be applied iteratively in the lifecycle to improve a user interface.
Since most ML models are developed to handle big data, a \QAforML interface is typically used in conjunction with a huge dataset.
With such an interface, users often have to spend much time performing laborious operations of viewing data objects and machine predictions and correcting errors when prediction errors are identified.
For such labor-intensive \QAforML processes, it is desirable to evaluate a diverse range of interface variants to identify the design option that maximizes the user's productivity.
Due to the labor-intensive nature of \QAforML, this paper focuses on the evaluation metric of \emph{task completion time} in \QAforML.

It is common to adopt a user-centered design process~\cite{Booshehrian2012Vismon,Wang2020Tac} to design a user interface following a requirements analysis phase.
After the software is developed, one conducts usability studies to evaluate the user interface.
Such user studies can inform the improvement of the user interface at different stages of the software lifecycle~\cite{Chau2011Apolo,Wu2018iTTVis}.
While opinions of potential users and usability studies can assist the design of such interfaces, these studies are usually costly and time-consuming as they intensively involve human subjects.
It is difficult to evaluate a large number of design options while taking into account various factors, such as users' skills and strategies in performing a large collection of operations.
It is also difficult to conduct user studies during the design phase, i.e., before a user interface has been developed.

This work proposes a model-based evaluation approach to efficiently evaluate \QAforML interfaces by modeling and simulating the task completion time.
We specifically focus on grid-based interfaces for \QAforML, which are common in the literature on data labeling systems.
Our approach is based on a model of the user's routine operations in \QAforML processes.
Our approach can be seen as an adaptation of the Keystroke-Level Model to estimate quality assurance tasks' completion time in grid-based interfaces~\cite{Card1980Keystroke}.

Using simulation, UI specialists (e.g., designers, developers, and evaluators) can explore design options at a scale unattainable with user studies.
We demonstrate applying this approach to optimize \QAforML interfaces in data \extract{}ion and aerial image classification systems.

We model and simulate the influence of various factors on the \QAforML time cost.
In this paper, we describe the model in a manner of increasing its complexity gradually.
In Section~\ref{sec:model-overview}, we describe our simulation approach, overview the simulation model, and outline the factors to be modeled and simulated.
We start with modeling simple \QAforML interfaces that only provide single edit commands.
Section~\ref{sec:factor-layout-application} focuses on the factors of interface layouts and application scenarios.
The simple modeling exercise demonstrates the feasibility of using simulation to guide the optimization of the interface layout.
The availability of interaction functions in \QAforML interfaces and algorithmic \module{}s may affect the time cost.
Batch edit functions enable users to edit the label of multiple data objects simultaneously.
Section~\ref{sec:factor-label-strategy} investigates the influence of introducing batch edit functions.
After adding the batch edit commands, users may accomplish \QAforML tasks with different combinations of single or batch edit commands.
In Section~\ref{sec:factor-label-strategy}, we introduce the factors about different user strategies in combining different commands.
The accuracy of the predictions made by the ML model is a factor that influences the overall \QAforML time cost. 
In Section~\ref{sec:factor-default-label-accuracy}, we investigate the modeling of the impact of this factor.
While \QAforML interfaces may randomly present data objects to users, intelligently ordering data objects can enable more uses of batch edit commands.
In Section~\ref{sec:factor-rank-method}, we bring into the simulation model the factor of algorithmic rank functions for ordering data objects.
In Section~\ref{sec:model-summary}, we summarize the overall simulation model with all the factors integrated.

The main contributions of this work include the following:
\begin{itemize}
    \item a simulation \emph{model} of users' routine operation sequence and operation time costs in quality assurance interfaces, and

    \item an \emph{application} of the model to evaluate and optimize quality assurance interfaces for data \extract{}ion and aerial image classification.
\end{itemize}

  \section{Related Work}

In the following, we review label quality assurance interfaces in interactive machine learning, which are the subjects of our evaluation method.
We provide a brief context of usability evaluation methods and then focus on the model-based evaluation approach to which our method belongs.

\subsection{Quality Assurance in Semi-automatic Labeling}

In interactive machine learning processes, the user may intervene various steps, such as feature selection~\cite{Alsallakh2014Visual}, model tuning~\cite{Chen2021Interactive}, testing~\cite{Chen2022HINT}, and explanation~\cite{Spinner2020explAIner}.
In quality-sensitive applications, quality assurance of model predictions and error identification is important~\cite{Khanna2022Finding}.
Among the various steps the user may intervene in, a common form of user involvement is to label data for iterative model training~\cite{Dudley2018Review}.
For example, interactive machine learning systems for information retrieval may utilize relevance feedback explicitly or implicitly labeled by the user to iteratively improve the relevance of the retrieved result~\cite{Huang2008Active}.
To reduce user burden, a common strategy is for the system to support semi-automatic labeling by learning from the user inputs and providing default labels.
This way, the user only needs to quality-assure the default labels and correct mistakes, which typically takes less time than labeling from scratch.

Typically, the default labels to be quality-assured in semi-automatic labeling are provided by machine-learned models.
Fluid Annotation~\cite{Andriluka2018Fluid} uses a pre-trained neural network to propose a set of possibly-overlapping segments to help annotate image segmentation.
The user can then edit the geometry of the segments to correct errors.
V-Awake~\cite{GarciaCaballero2019V} focuses on time series segmentation.
It uses LSTM to assign tentative labels and visualizes the model information to help annotators diagnose and correct model predictions.

In some cases, the default labels to be quality-assured may also come from crowdworkers.
LabelInspect~\cite{Liu2019Interactive} is a visual analytics system for improving crowdsourced image labels.
It enables the user to identify uncertain instance labels and unreliable crowdworkers.
DataDebugger~\cite{Xiang2019Interactive} is a system for image label correction based on data debugging using trusted items.
It allows the user to identify and correct suspicious data labels, specify data labels to be trusted, and propagate the trusted labels.

When the default labels come from machine predictions, the machine assistance may be gradually improved by incrementally learning from user inputs to reduce users' quality assurance effort.
Fails and Olsen propose Crayons~\cite{Fails2003Interactive} for image pixel classification.
The user can paint pixels to correct machine predictions and simultaneously provide training data to the machine.
Fogarty et al. develop a search system, CueFlik~\cite{Fogarty2008CueFlik}, with which the user can define search criteria for a concept by providing positive and negative examples and rank search results according to the similarity to the concept.
ISSE~\cite{Bryan2014ISSE} algorithmically suggests refined segmentation to help annotators separate sound into its respective sources by painting on time-frequency visualizations.

\subsection{Usability Evaluation}

Evaluation is helpful in various stages of iterative software engineering processes for interactive systems.
During prototyping, evaluation may guide software development to improve usability.
After implementation, evaluation may derive summative results on whether the interactive system achieves specific goals, such as improved efficiency over existing systems.

\subsubsection{Overview}

Usability evaluation methods can be categorized by two dimensions: involvement of users and methodological approach~\cite{Barkhuus2007Mice}.
An empirical evaluation involves a sample of target users (also referred to as ``testing method''~\cite{Andrews2008Evaluation}).
By comparison, an analytical evaluation usually requires evaluation experts to carry out (also referred to as ``inspection method''~\cite{Andrews2008Evaluation} and ``expert analysis''~\cite{Dix2004Human}).
The evaluation methodology can be qualitative, quantitative, or combined.

Empirical evaluation captures payoff measures, such as user performance, to infer interface properties~\cite{Gray1998Damaged} with the following example methods.
\begin{itemize}
    \item \emph{Laboratory studies} involve a sample of users to perform tasks and analyze objective measurements, such as task completion time and accuracy.
        It is prevalent in HCI research nowadays.

    \item \emph{Think-aloud} gains insight about the interface through the records of what the user said that came into their mind while performing tasks with the interface.

    \item \emph{Questionnaires}, such as Likert scales, gather user ratings of interfaces.
\end{itemize}

Analytical evaluation inspects the interface's intrinsic properties influencing usability~\cite{Gray1998Damaged} with the following methods being examples.
\begin{itemize}
    \item \emph{Heuristic evaluation} involves a team of evaluators checking an interface against a short list of design guidelines and aggregating the identified issues~\cite{Nielsen1990Heuristic}.
    
    \item \emph{Cognitive walkthroughs} are a technique where evaluators mentally walk through a task following a novice user's mindset to identify usability issues~\cite{Lewis1990Testing}.

    \item \emph{Model-based evaluation} involves modeling users' physical and mental operations to accomplish a task in an interface~\cite{Sears2007Human}.
        It is also referred to as ``action analysis''~\cite{Lewis1993Task}.
        The modeling enables the estimation of usability metrics, such as task completion time~\cite{Card1980Keystroke} and learning time costs~\cite{Kieras1988Towards}.
\end{itemize}

Analytical methods are usually carried out by evaluation experts.
Analytical methods do not directly measure the performance of target users and thus may not be as conclusive as empirical methods.
Meanwhile, analytical methods can be less costly than empirical methods, as the former typically involve fewer people and thus are more suited for early-stage design and prototyping iterations.
Additionally, it is usually easier to attribute the causes of usability issues with analytical methods.
Our approach belongs to analytical methods, specifically model-based evaluation.

\subsubsection{Model-based Evaluation}

We use the model-based evaluation approach to evaluate quality assurance interfaces in interactive machine learning.
Generally, model-based evaluation involves modeling the user's task completion procedure in an interactive system to estimate usability metrics, such as task completion time or learning time.
Kieras identifies three categories of models in model-based evaluation: task network models, cognitive architecture models, and GOMS models~\cite{Sears2007Human}.
We introduce examples following this categorization.

\begin{itemize}
    \item \emph{Task network models} capture the dependencies between user and machine tasks as networks~\cite{Kraiss1981Display}.
    SAINT (Systems Analysis of Integrated Networks of Tasks) is a simulation language developed for task network models~\cite{Chubb1981SAINT}.
    Given the distributions of probabilistic variables in the model, it supports the estimation of task completion time distribution through Monte Carlo experiments.

    \item \emph{Cognitive architecture models} apply theories in cognitive science to model human motor control, perception, and cognition~\cite{Anderson1983Architecture}.
    Examples include the ACT-R architecture~\cite{Anderson1993Rules} and the EPIC architecture~\cite{Kieras1997Predictive}.
    Both of them model users' low-level interaction with production rules that depict user reactions to stimuli but differ in some aspects of their human model.
    ACT-R concerns serial production cycles, while EPIC concerns parallel production cycles.

    \item \emph{GOMS models} describe the required procedural knowledge for a user to carry out routine tasks in user interfaces, similar to depicting computer algorithms with pseudocode~\cite{John1996GOMS,John1996Using}.
    The term ``GOMS'' stands for its four components: goals, operators, methods, and selection rules.
    Various GOMS techniques exist, such as the Keystroke-Level Model~\cite{Card1980Keystroke}.
    Our work falls into this category.
\end{itemize}

Model-based evaluation has been used in various application scenarios.
The EPIC architecture has been used to predict human performance in telephone operator tasks~\cite{Kieras1997Predictive}.
John and Vera use GOMS to model the highly interactive task of video game playing~\cite{John1992GOMS}.
Gond and Kieras use GOMS to identify usability issues and aid the redesign of ergonomics assessment software~\cite{Gong1994Validation}.
Beard et al. apply GOMS to evaluate the user interface for displaying computed tomography images for medical diagnosis~\cite {Beard1996Quick}.
Kieras and Santoro apply GOMS to evaluate the user interface and collaboration scheme for shipboard workstations~\cite{Kieras2004Computational}.
Kieras and Hornof introduce a GOMS model for the visual search task~\cite{Kieras2014Towards}.
Ramkumar et al. apply GOMS analysis to image segmentation interfaces~\cite{Ramkumar2017Using}.
Azzopardi et al. model and simulate the trade-off between the query and assess interactions to the total time cost in information retrieval systems~\cite{Azzopardi2011Economics}, and validate the simulation result through a user study~\cite{Azzopardi2013How}.
Like this line of work, our modeling and simulation follow the model-based evaluation approach.
We focus on a different scenario of evaluating data \extract{}ion and aerial image classification systems.

Task completion time is a common metric in usability evaluation.
To estimate task completion time, the model-based evaluation typically involves the modeling or reusing existing models of user operations' time costs.
For example, Fitts' law~\cite{Fitts1992Information} has long been used to model the time costs of users' object-pointing operations~\cite{MacKenzie1992Extending}.
It has been extended to the time cost modeling in related mouse movement tasks, such as trajectory-based tasks~\cite{Accot1997Fitts} and object-pointing tasks with multi-scale navigation~\cite{Guiard2004View}.

With models in model-based evaluation, redesign and optimization of user interfaces can be conducted computationally and swiftly, as demonstrated in the computational interaction literature~\cite{Oulasvirta2018Computational}.
Gajos and Weld propose the SUPPLE system for automatic user interface generation through constrained optimization of users' navigation costs~\cite{Gajos2004SUPPLE}.
The optimization is based on the interface model with functional specifications, the model of the rendering device, and the users' usage trace in the interface.
Zhang and Zhai propose a card-playing model for information retrieval interfaces and demonstrate adaptive optimization of the interface according to the screen size and user's interaction sequence~\cite{Zhang2015Information}.
Todi et al. propose an adaptive user interface technique that uses Fitts' law to estimate user interaction cost and reinforcement learning to plan sequences of adaptions~\cite{Todi2021Adapting}.

Like this line of research, we demonstrate using model-based evaluation to optimize interfaces, with estimated task completion time being the metric.
Meanwhile, our model is dedicated to the scenario of quality assurance in interactive machine learning.
While evaluating interactive machine learning systems is generally challenging~\cite{Boukhelifa2020Challenges}, our work suggests that some of the most laborious routine processes in interactive machine learning, such as quality assurance, can be modeled effectively and evaluated efficiently.

A limitation of model-based evaluation methods is their restriction to routine and well-defined user tasks~\cite{John1996GOMS}.
For exploratory tasks where the goal is ill-defined and requires users' complex reasoning procedures and creativity, it is intractable for the evaluators to model the usage procedures.
This problem is absent in the case of quality assurance that we focus on, which has well-defined goals and relatively predictable usage procedures.
Thus, we can derive a model of the user operation sequences.

  \section{Background}

This section introduces two example data-centric applications and their \QAforML interfaces to illustrate the need for \QAforML.
Then, we present a template grid-based interface that typifies the \QAforML interfaces.

\subsection{Data-centric Applications}
\label{sec:application}

\begin{figure}[htbp]
    \centering
    \includegraphics[width=\linewidth]{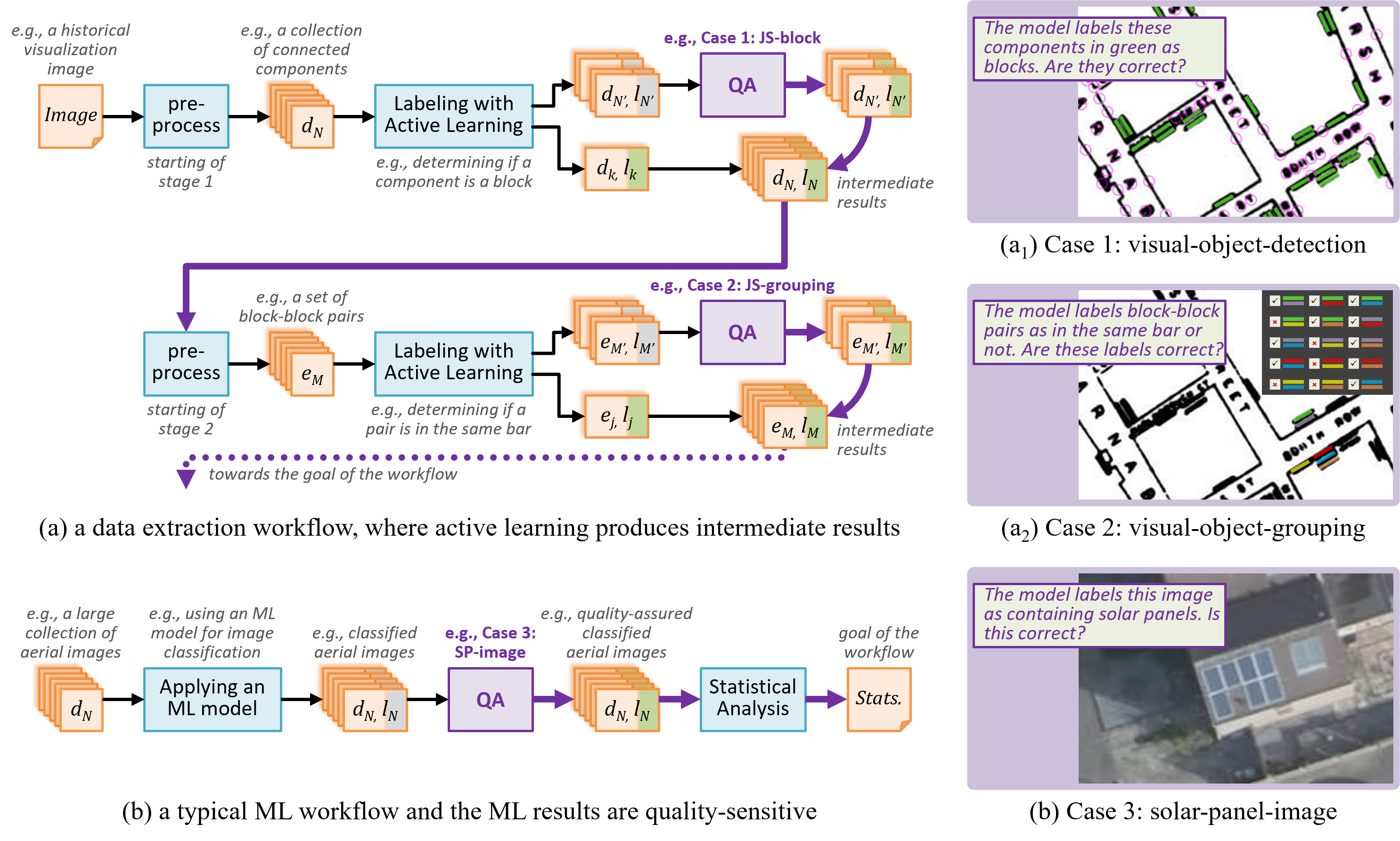}
    \caption{
        ML \workflow{}s in data-centric applications requiring quality assurance:
        (a) A \workflow{} with two stages ($\mathrm{a_1}$ and $\mathrm{a_2}$) to \extract{} datasets from visualization images.
        Quality assurance is necessary to avoid propagating errors to the subsequent stage and guarantee the correctness of the \extract{}ed dataset.
        (b) A \workflow{} to generate summary statistics from the classification of whether an image contains solar panels.
        Quality assurance of classification labels is needed to achieve reliable statistics.
    }
    \Description[
        Fully described in the main text of the section.
    ]{}
    \label{fig:application-workflows}
\end{figure}

\textbf{Visualization Data \Extract{}ion:}
The first application concerns detecting and grouping visual objects in historical visualizations as introduced by Zhang et al.~\cite{Zhang2021MI3}.
For example, given John Snow's famous cholera map~\cite{Snow1855Mode}, the rectangular blocks are to be detected (Figure~\ref{fig:application-workflows}($\mathrm{a_1}$)) and grouped (Figure~\ref{fig:application-workflows}($\mathrm{a_2}$)).
The detected visual objects are then used to \extract{} the underlying datasets generating the visualizations.
While machine-learned models may not accurately detect the visual objects, they can suggest visual objects and tentative classification labels for the user to quality-assure, as demonstrated in the MI3 \workflow{}s~\cite{Zhang2021MI3}.
In this application, QA is needed for the correctness of the \extract{}ed datasets.

\textbf{Solar Panel Detection:}
The second application concerns generating a labeled benchmark image dataset where labels classify aerial images as containing solar panels or not (Figure~\ref{fig:application-workflows}(b)).
The benchmark dataset is used to generate official statistics on solar panel deployment.
In this application, while using ML models' predictions may accelerate the labeling, predictions not checked manually are imprecise and untrustworthy.
A QA process inspecting and correcting machine predictions is necessary for the correctness of the official statistics.
Figure~\ref{fig:application-workflows}(b) shows this process of applying an ML model and quality-assuring machine predictions to generate a benchmark dataset and, consequently, official statistics.

These applications typify scenarios of carrying out real-world \emph{data-centric} ML projects.
In these scenarios, the ultimate goal is to gather high-quality labeled/processed data.
Using ML in these projects is not to train accurate models but to assist the data labeling and processing.
These applications exhibit a common challenge: a user may need to perform many tedious quality assurance operations to correct machine prediction errors.
It is critical to optimize \QAforML interfaces to reduce user operation costs.

\subsection{Grid-based Interfaces}
\label{sec:grid-based-interface}

\begin{figure}[!htbp]
    \centering
    \includegraphics[width=\linewidth]{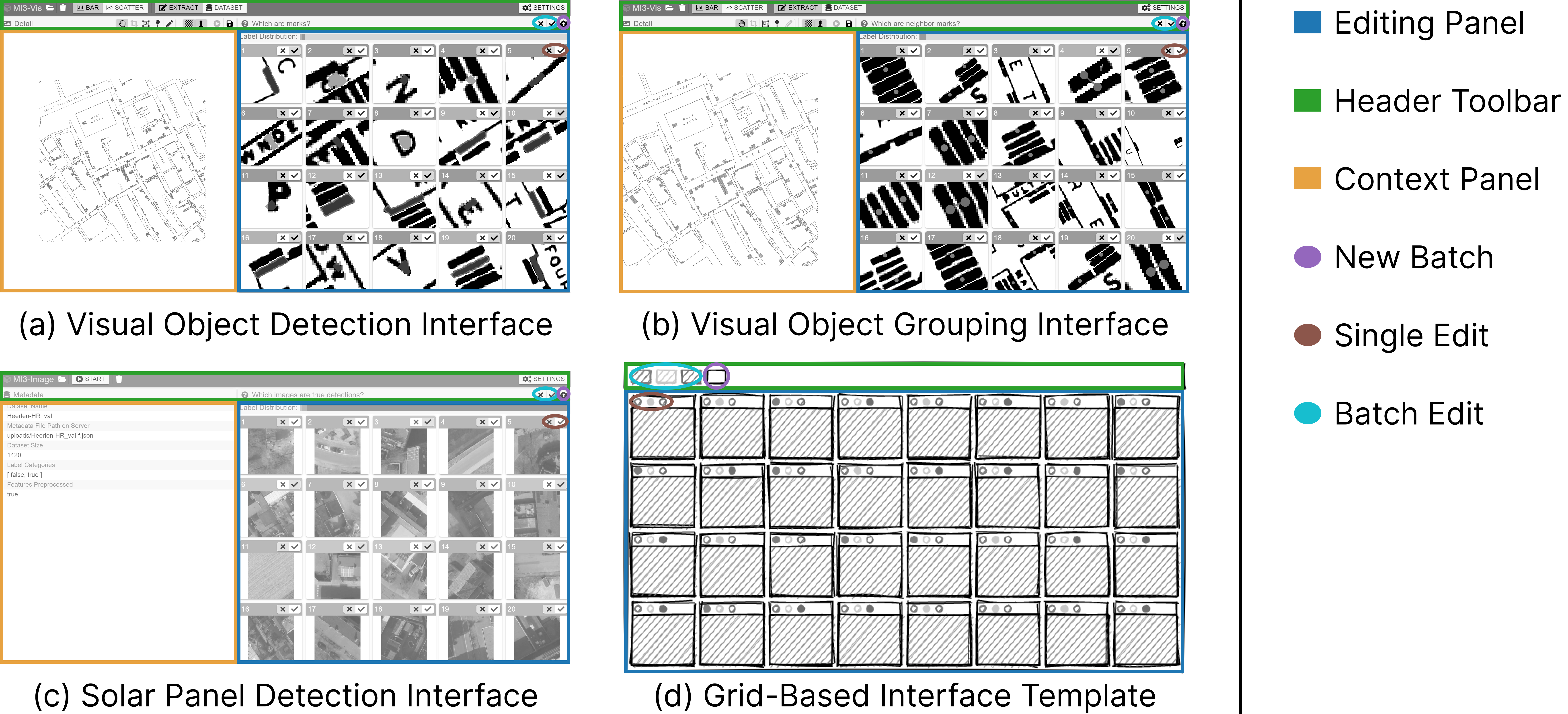}
    \caption{
        Examples of \QAforML interfaces:
        The interface components (editing panel, header toolbar, and context panel) are annotated with colored rectangles.
        The buttons corresponding to operations (new batch, single edit, batch edit) available to the user are annotated with colored ovals.
        The interfaces all contain an edit panel that presents a matrix of thumbnails showing data objects, a header toolbar providing interaction functions, and a context panel that provides an overview of the dataset:
        (a) An interface for visual object detection corresponding to \workflow{} in Figure~\ref{fig:application-workflows}($\mathrm{a_1}$).
        (b) An interface for visual object grouping corresponding to \workflow{} in Figure~\ref{fig:application-workflows}($\mathrm{a_2}$).
        (c) An interface for solar panel detection corresponding to \workflow{} in Figure~\ref{fig:application-workflows}(b).
        (d) The typical design of grid-based interfaces in \QAforML.
    }
    \Description[
        Fully described in the caption.
    ]{}
    \label{fig:grid-based-interfaces}
\end{figure}

\QAforML requires users to verify machine predictions in interfaces.
Figure~\ref{fig:grid-based-interfaces} shows three \QAforML interfaces corresponding to the two applications described above.
Figure~\ref{fig:grid-based-interfaces}(a, b) show interfaces for classification in visualization data \extract{}ion.
The interface in Figure~\ref{fig:grid-based-interfaces}(a) displays a batch of visual objects.
The visual objects are to be categorized as encoding data or not.
The context panel on the left shows part of John Snow's cholera map under check.
Figure~\ref{fig:grid-based-interfaces}(c) shows an interface for classifying solar panel detection.
In this interface, an ML model assigns tentative classification labels for the images on whether they contain solar panels.
In these interfaces, ML models are used to extract and group visual objects.

The quality assurance interface can be divided into three parts: the editing panel, header toolbar, and context panel.
The editing panel is where a user performs most operations to correct ML prediction errors.
Because data objects to be inspected are usually too many to be quality-assured all at once, the editing panel normally shows a subset (referred to as a ``batch'') of data objects each time.
The header toolbar may provide utilities such as dataset upload, requesting the system to present the next batch of data objects to be quality-assured, and simultaneously setting labels for a batch of data objects.
For example, in Figure~\ref{fig:grid-based-interfaces}(a - c), each interface shows a batch of 20 data objects.
The context panel gives the user a dataset overview and may present summary statistics about the labeling progress, such as the machine errors identified.
The following mainly focuses on the editing panel and header toolbar as they accommodate the functions directly related to quality assurance.

In all the interfaces, users need to manually verify machine predictions in a grid panel on the right part of the interface.
The grid panel presents each data object as a thumbnail image in a grid cell.
Each grid cell provides buttons for editing its label.
Figure~\ref{fig:grid-based-interfaces}(d) typifies the design of grid-based interfaces.
This paper focuses on grid-based interfaces in \QAforML.
We refer to each thumbnail corresponding to a data object and an interaction mechanism (e.g., buttons for editing label categories) as a \emph{grid cell}, and all grid cells in the editing panel as a \emph{batch}.
It is the most commonly adopted design in \QAforML interfaces according to Zhang et al.'s survey~\cite{Zhang2022OneLabeler} with numerous instances in the literature~\cite{Liu2019Interactive,Xiang2019Interactive,Baeuerle2020Classifier,Hoeferlin2012Inter,Zhang2021MI3}.

We list typical functionalities of \QAforML interfaces exhibited in the literature.
The interface enables a user to \emph{request a new batch} of data objects to be quality-assured in the editing panel.
An algorithmic process (e.g., active learning or clustering) may determine the priority for data objects to be presented to the user.
An ML model assigns default labels to the data objects.
The user has to inspect data objects to judge whether the default labels are correct.
The interface provides \emph{single edit} commands in the editing panel for correcting individual label errors.
To improve efficiency, the interface provides the user with \emph{batch edit} commands for changing the labels of a group of data objects (e.g., set all labels to positive).

\QAforML interfaces utilize human intelligence to

\begin{itemize}
    \item recognize ML model's prediction errors, and
    \item determine whether using a batch edit command is more efficient than applying single edit commands to data objects individually.
\end{itemize}

Batch editing is beneficial when the batch of data objects mostly shares the same label.
Consider a batch of 20 data objects with binary labels (e.g., ``yes'' or ``no'').
Suppose there are 10 true-positive labels, 2 false-positive labels, and 8 false-negative labels.
In this case, the user can use a batch edit command to set all labels to positives first and then change the 2 false-positive labels to negatives.
This process requires 3 user operations.
If the user only uses single edit commands, 10 operations are needed.

\QAforML interfaces also utilize machine assistance to

\begin{itemize}
    \item assign default labels to data objects, and
    \item determine the ordering for data objects to be labeled.
\end{itemize}

Suitable ordering may accelerate the quality assurance process.
Firstly, the ordering algorithm may bunch together data objects of the same label, providing more opportunities for applying a batch edit command.
Secondly, when the ML model for assigning default labels is incrementally updated with the user corrections, ordering generated by active learning methods may enable the ML model to learn more efficiently, leading to better default labels.

While human-in-the-loop is indispensable for \QAforML processes, it is also essential to evaluate and optimize \QAforML interfaces to reduce user effort.
Meanwhile, evaluating different design options of \QAforML interfaces with user-centered evaluation methods (e.g., surveys, group discussions, controlled experiments) is usually time-consuming for designers and potential users participating in the evaluation.

\section{Overview of the Simulation Approach}
\label{sec:model-overview}

This section outlines a model-based approach to evaluating quality assurance interfaces.
Section~\ref{sec:workflow} introduces a general \workflow{} capturing the user's routine operations in grid-based interfaces for \QAforML.
We describe \module{}s in the \workflow{} to be modeled and simulated.
The \workflow{} is constructed by analyzing user and machine operations in the \QAforML interfaces typified in Figure~\ref{fig:grid-based-interfaces}(d).
Given the \workflow{}, Section~\ref{sec:simulation} introduces a simulation model.
The model estimates the time cost of a \QAforML session by simulating sequences of user and machine operations and adding up the time cost of each operation.
Section~\ref{sec:factors} outlines the factors we model and simulate.
Section~\ref{sec:scope} clarifies the scope and assumptions of the simulation approach.

\subsection{A Routine Task \Workflow{}}
\label{sec:workflow}

\subsubsection{Problem Formulation}

In general, quality assurance concerns the following setup.
A dataset $X = \{x_i | i = 1, 2, \ldots, n\}$ contains $n$ data objects to be assigned quality-assured labels.
An ML model is applied to the dataset $X$, generating default labels $Y = \{y_i | i = 1, 2, \ldots, n\}$ where $y_i$ is the default label for $x_i$.
The user needs to quality-assure the default labels to generate verified labels $Y^* = \{y_i^* | i = 1, 2, \ldots, n\}$.

We refer to the entire process of carrying out operations to quality-assure the $n$ data objects as a \emph{session}.
In typical \QAforML interfaces, a subset of data objects, $X_i \subseteq X$, is loaded into the editing panel.
The editing panel can accommodate maximal $n_{batch}$ data objects, and a subset of data objects loaded to the editing panel is of size $K$ ($K \leq n_{batch}$).
We refer to the process of quality-assuring the subset $X_i$ as a \emph{round}.

For classification tasks, let $c$ be the number of label categories.
Let $cm \in [0, 1]_{c \times c}$ be the normalized confusion matrix with $\sum_{i, j} cm_{i, j} = 1$.
$cm_{i, j}$ denotes the rate of the data objects with true label $i$ and default label $j$ over all the $n$ data objects.

For binary classification, the correctness of each label $y_i$ can be represented with one of the four values: \textbf{true-positive (TP)}, \textbf{true-negative (TN)}, \textbf{false-positive (FP)}, and \textbf{false-negative (FN)}.
Let $n_{TP}$, $n_{TN}$, $n_{FP}$, $n_{FN}$ be the number of TPs, TNs, FPs, and FNs in the whole dataset.
The normalized confusion matrix can be represented as
$cm = \frac{1}{n}
    \big(\begin{smallmatrix}
            n_{TN} & n_{FP}\\
            n_{FN} & n_{TP}
        \end{smallmatrix}\big)
$.

\subsubsection{Operators in the \Workflow{}}

Section~\ref{sec:grid-based-interface} introduces user operations and machine operations in typical grid-based interfaces for \QAforML.
Using the user and machine operations as building blocks, we summarize the quality assurance \workflow{} as shown in Figure~\ref{fig:batch-edit-workflow}.
The \workflow{} includes two nested loops.
The inner loop is between the operations ``view an object'' and ``batch ends?''.
Going through the inner loop once quality-assures the label of a data object.
The outer loop is between the operations ``sorting remaining'' and ``session ends?''.
Going through the outer loop once quality-assures the labels of a batch of data objects.
In the following, we introduce each operation in the \workflow{}.

\begin{figure*}[!htbp]
    \centering
    \includegraphics[width=\linewidth]{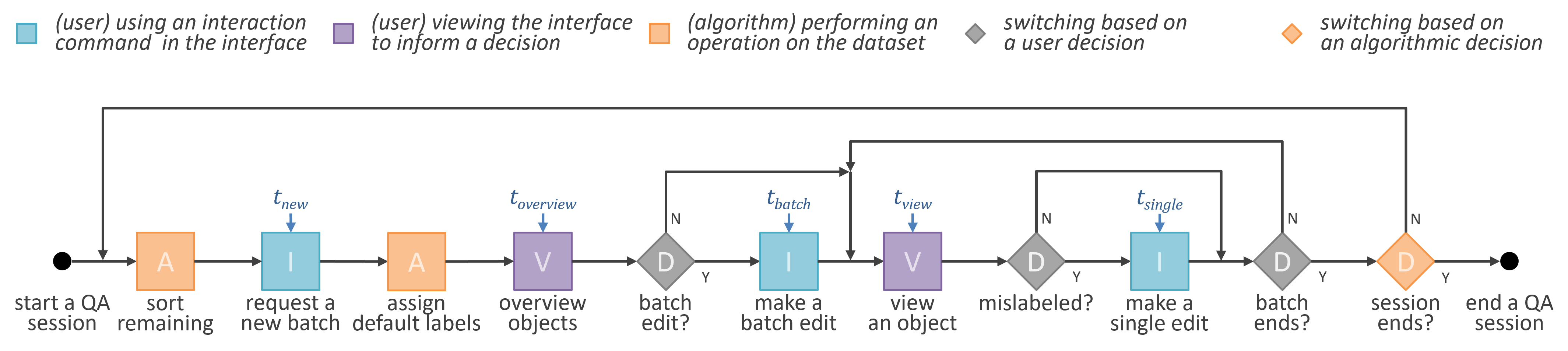}
    \caption{
        A general \workflow{} of user and machine operations in a quality assurance session in a grid-based interface typified in Figure~\ref{fig:grid-based-interfaces}(d).
        We group the operations into five categories: user interaction in the interface, the user's viewing action, algorithmic operation, user decision, and algorithmic decision.
    }
    \Description[
        Fully described in the main text of the section.
    ]{}
    \label{fig:batch-edit-workflow}
\end{figure*}

The \workflow{} involves a set of interaction commands, or operators, initiated by the user (shown as blue rectangles in Figure~\ref{fig:batch-edit-workflow}), including:

\begin{itemize}
    \item \emph{Request a new batch} of data objects to quality-assure by activating a ``new batch'' command in the interface (denote its average time cost as $t_{new}$).
    \item \emph{Make a batch edit} to assign the same label to the current batch of data objects by activating a ``batch edit'' command in the interface (time cost $t_{batch}$).
    \item \emph{Make a single edit} to edit the label of a data object in the current batch by activating a ``single edit'' command in the interface (time cost $t_{single}$).
\end{itemize}

Some of the interaction commands can only be issued after the user makes explicit decisions (shown as gray diamonds in Figure~\ref{fig:batch-edit-workflow}), including:

\begin{itemize}
    \item \emph{Batch Edit?} The decision of whether a ``batch edit'' should be issued, which depends on whether using a batch edit saves effort over only using single edits.
    \item \emph{Mislabeled?} The decision of whether a ``single edit'' should be issued, which depends on whether the data object is mislabeled.
    \item \emph{Batch Ends?} The decision of whether all the data objects in the current batch are quality-assured.
\end{itemize}

The user needs to perform viewing actions (shown as purple rectangles in Figure~\ref{fig:batch-edit-workflow}) to obtain the information of the interface state to inform some of the decisions, including:

\begin{itemize}
    \item \emph{Overview objects} in the current batch to comprehend the overall label category distribution (time cost $t_{overview}$).
    \item \emph{View an object} in the current batch to comprehend its label category and whether it is mislabeled (time cost $t_{view}$).
\end{itemize}

The machine needs to perform algorithmic operations, including:

\begin{itemize}
    \item \emph{Sort remaining} to reorder remaining data objects to be quality-assured.
          In a \QAforML interface, the sorting can be implemented with any algorithm that serves the reordering purpose.
          For example, the rank method can be implemented with a clustering algorithm to group data objects of the same label category~\cite{Tang2013Towards,Cui2007EasyAlbum,Suh2007Semi}.
          In this case, the rank method may create good opportunities to use batch edit commands.
          The sorting may also be implemented by scoring functions in active learning~\cite{Settles2008Active} to prioritize data objects whose labels are more uncertain.
    \item \emph{Assign default labels} to the batch of selected data objects.
          In the scenario of quality-assuring ML predictions, default labels may be assigned by a pre-trained machine learning model or an incrementally updated model~\cite{Bryan2014ISSE,Andriluka2018Fluid,GarciaCaballero2019V}.
          In general quality assurance scenarios, default labels may come from other sources, such as another user or pre-defined rules.
\end{itemize}

An algorithmic decision (\emph{session ends?}) determines whether the session should end, which depends on whether all the data objects are quality-assured (shown as an orange diamond in Figure~\ref{fig:batch-edit-workflow}).

The original Keystroke-Level Model decomposes the interaction tasks into the keystroke-level operators, such as keystrokes, pointing, and mental preparation for physical actions, and assumes each type of operator costs the same amount of time.
To estimate the task completion time more accurately, we decide not to make this assumption.
For example, we do not assume the time cost of ``batch edit'' ($t_{batch}$) and ``single edit'' ($t_{single}$) to be the same, considering that the buttons for batch edit and single edit are distributed at different places in the interface.
By comparison, in the Keystroke-Level Model, ``batch edit'' and ``single edit'' are both a pointing followed by a clicking, and thus should be assumed to cost the same amount of time.

\subsection{A Simulation Model}
\label{sec:simulation}

The \workflow{} provides the basis of a simulation model and defines the top-level structure of a simulator implementing the model.
Algorithm~\ref{alg:session-simulation} illustrates an implementation of a simulator for \QAforML sessions.
The simulation algorithm generates a mock dataset, synthesizes a sequence of operations to quality-assure the mock dataset, and adds up the operation time costs.

\renewcommand{\algorithmicrequire}{\textbf{Input:}}
\renewcommand{\algorithmicensure}{\textbf{Output:}}
\algdef{SE}[DOWHILE]{Do}{DoWhile}{\algorithmicdo}[1]{\algorithmicwhile\ #1}%

\begin{algorithm}[htbp]
    \small
    \caption{\QAforML Session Simulation}\label{alg:session-simulation}
    \begin{algorithmic}[1] %
        \Require $ n $, $ n_{batch} $, $\vec{t}$
        \Ensure time cost of the quality assurance session $ T_{session} $
        
        \State $ \mathrm{X, Y^* \gets CreateMockDataset(n)} $
        \State $ \mathrm{ops \gets []} $ \Comment{log of the operation sequence}
        \Do
            \State $ \mathrm{X, Y \gets Rank(X, Y)} $ \Comment{sorting remaining}
            \State ops.push(``new'') \Comment{request a new batch}
            \State $ \mathrm{X_s \gets X.slice(n_{batch})} $
            \State $ \mathrm{Y^*_s \gets Y^*.slice(n_{batch})} $
            \State $ \mathrm{Y_s \gets AssignDefaultLabels(X_s)} $ \Comment{assign default labels}
            \State ops.push(``overview'') \Comment{overview objects}
            \State $ \mathrm{cmd \gets SelectEditCommand(Y_s, Y^*_s)} $
            \If{$ \mathrm{cmd \in BatchEditCommands} $} \Comment{batch edit?}
                \State ops.push(cmd) \Comment{make a batch edit}
                \State $ \mathrm{Y_s \gets ApplyBatchEdit(cmd, Y_s)} $
            \EndIf
            \For{$ \mathrm{j \in (1, ..., Y_s.length)} $} \Comment{loop until batch ends}
                \State ops.push(``view'') \Comment{view an object}
                \If{$ \mathrm{Y_s[j] \ne Y^*_s[j]} $} \Comment{mislabeled?}
                    \State ops.push(``single'') \Comment{make a single edit}
                \EndIf
            \EndFor
        \DoWhile{$ \mathrm{Y \ne []} $} \Comment{session ends?}
        \State $ \mathrm{T_{session} \gets} $ GetCost(ops, $ \vec{\mathrm{t}} $) \Comment{compute total time cost}
        \State \Return{$ \mathrm{T_{session}} $}
    \end{algorithmic}
\end{algorithm}

The simulation algorithm takes as input the number of data objects $n$, the number of data objects presented simultaneously in the editing panel $n_{batch}$, and time costs of the operations in the \workflow{} $\vec{t}$.
It outputs the total time cost of the QA session.

Algorithm~\ref{alg:session-simulation} is generic in that different implementations of routines such as ``CreateMockDataset'', ``Rank'' and ``AssignDefaultLabels'' can be plugged in to simulate different \QAforML interface designs.
For example, at the beginning of the simulation (line 1 of Algorithm~\ref{alg:session-simulation}), a mock dataset is generated by ``CreateMockDataset''.
This routine may fetch real data objects as is in the simulations in Section~\ref{sec:factor-layout-application}.
It may alternatively generate a mock dataset according to $n$ and an additional parameter of the default labels' confusion matrix, as is in the simulations in Sections~\ref{sec:factor-label-strategy}, \ref{sec:factor-default-label-accuracy}, and \ref{sec:factor-rank-method}.

\subsection{Modeled and Simulated Factors}
\label{sec:factors}

This section outlines factors of \QAforML interfaces that may affect the total time cost of the quality assurance process.
These factors inform our modeling and simulations.

\begin{itemize}
    \item \textbf{Interface layout:}
          The grid-based interface may utilize different layouts (as shown in Figure~\ref{fig:interface-layouts}).
          The display size of each data object is affected by the layout, assuming the screen size is fixed.
          The display size may affect the time cost to comprehend the data object's label ($t_{view}$ and $t_{overview}$).
          Additionally, the layout determines the number of data objects displayed simultaneously ($n_{batch}$), affecting the effectiveness of batch edit commands.
          For layouts with larger $n_{batch}$, issuing a batch edit command can edit the labels of more data objects.
          Section~\ref{sec:factor-layout-application} discusses this factor.
    \item \textbf{Application scenario:}
          The user operation time costs may depend on the quality assurance process's application scenarios (e.g., the ones introduced in Section~\ref{sec:application}).
          For example, the complexity and time cost of comprehending the data labels ($t_{view}$ and $t_{overview}$) may depend on the application scenario.
          Section~\ref{sec:factor-layout-application} discusses this factor.
          We model user operation time costs as functions of the \emph{layout} of the grid-based interface and the \emph{application} from which the dataset to be quality-assured comes.
    \item \textbf{Availability of interface functions:}
          Section~\ref{sec:workflow} introduces a general \QAforML \workflow{} that concerns several functions commonly but not always offered in \QAforML interfaces, such as batch editing and the rank method.
          The availability of these functions may affect the time cost of the \QAforML process.
          Section~\ref{sec:factor-label-strategy} discusses the influence of introducing batch edit commands.
          Section~\ref{sec:factor-rank-method} discusses the influence of introducing rank methods.
    \item \textbf{User's label strategy:}
          As the \QAforML interface provides more functionalities, there may be multiple methods to accomplish the same goal.
          The user's strategy to select a method for accomplishing the goal affects the time cost.
          Section~\ref{sec:factor-label-strategy} discusses this factor.
    \item \textbf{Default label accuracy:}
          The accuracy of default labels affects the number of times the user needs to use single edit and batch edit commands.
          Section~\ref{sec:factor-default-label-accuracy} discusses this factor.
    \item \textbf{Rank method:}
          The rank method may change the distribution of label categories in each batch.
          Effective rank methods may alter the distribution to create more opportunities for batch edit commands to label multiple data objects with one click.
          Section~\ref{sec:factor-rank-method} discusses this factor.
\end{itemize}

\subsection{Scope and Assumptions}
\label{sec:scope}

In the following sections, we set the following restrictions on the scope of our simulation-based evaluation:

\begin{itemize}
    \item We focus on the scenario of \emph{quality assurance of classification labels}.
          The following sections analyze the quality assurance of binary classification labels, with the two label categories being ``positive'' and ``negative''.
          Our analysis can easily be extrapolated to multi-class classification.

    \item We focus on the \emph{task completion time} as the evaluation metric.
\end{itemize}

\noindent We make the following assumptions to simplify the modeling and simulations:

\begin{itemize}
    \item We assume \emph{error-free execution} in the user operations.
          The user is assumed to be able to recognize the correct labels of the data objects and does not accidentally activate incorrect commands when using \QAforML interfaces.
          Fully capturing the user strategies and potential mistakes is intractable due to the current insufficient understanding of the human decision process.
          Thus, a general strategy in model-based evaluation methods is to model what the users should do instead of what the users will do~\cite{Sears2007Human}.
          We essentially give a lower-bound estimation by adopting the original assumption in the Keystroke-Level Model that the user makes no mistakes.

    \item We assume the \emph{machine operations are fast} and their time costs are ignorable compared with user operation time costs.
          Thus, we focus on the time costs of user operations.

    \item We assume the \emph{user operations are sequential} instead of parallel.
          Under this assumption, $GetCost$ in Algorithm~\ref{alg:session-simulation} is a simple summation.
          $GetCost(ops, \vec{t}) = \sum_{op} N_{op}t_{op}$ where $op$ denotes a user operation type, $N_{op}$ denotes the number of operations of type $op$, and $t_{op}$ denotes the time cost to execute the operation of type $op$ once.
\end{itemize}

The following sections introduce different technical aspects of our simulation model.
We gradually bring out factors to be modeled, increasing the model's complexity.

  \section{The Factors of Interface Layout and Application}
\label{sec:factor-layout-application}

The following introduces simulations of task completion time with different interface layouts and application scenarios.
We introduce a method to estimate user operations' time costs in the grid panel for editing labels.
We show how simulation can help find an optimal layout for the grid-based interface.
This section considers a simplified \QAforML interface template compared to Section~\ref{sec:grid-based-interface}.
This simplified interface template provides the new batch and single edit commands but does not provide batch edit commands.
Figure~\ref{fig:single-edit-workflow} shows its \workflow{}.

\begin{figure*}[!htbp]
    \centering
    \includegraphics[width=0.7\linewidth]{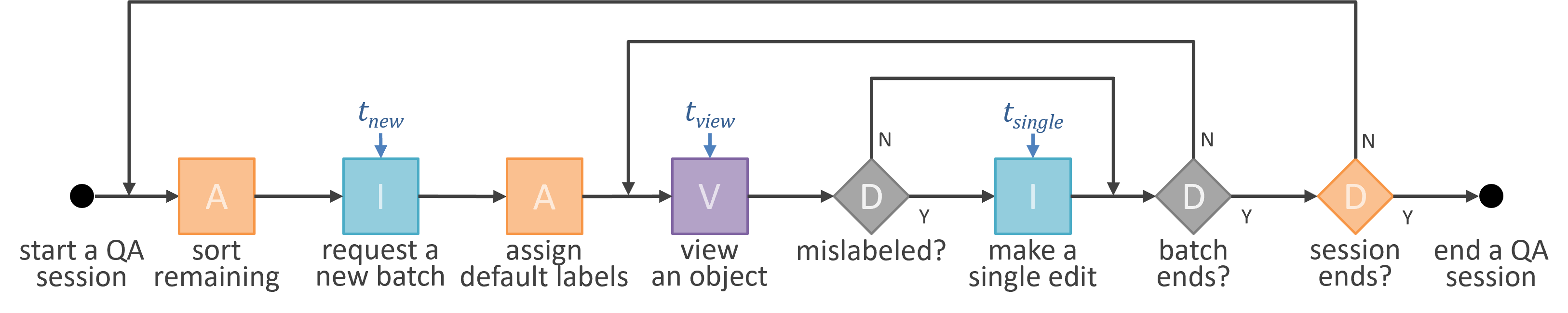}
    \caption{
        A \workflow{} of human and machine operations in a grid-based interface without batch edit functions.
        This \workflow{} simplifies the \workflow{} in Figure~\ref{fig:batch-edit-workflow} by removing operations related to batch edit.
    }
    \Description[
        The figure shows a workflow of human and machine operations.
        The workflow has the following steps: ``start a QA session'', ``sort remaining'', ``request a new batch'', ``assign default label'', ``view an object'', ``mislabeled?'', ``make a single edit'', ``batch ends?'', ``session ends?'', and ``end a QA session''.
        Some of the steps are associated with conditional branching.
        At the ``mislabeled?'' step, if the human decides that the data object is not mislabeled, the ``make a single edit'' step can be skipped.
        At the ``batch ends?'' step, if the human decides that there remain data objects that have not been quality-assured in the current batch, the next step is not ``session ends?'' but instead goes back to ``view an object''.
        Similarly, at the ``session ends?'' step, if the human decides that there remain data objects that have not been quality-assured in the current session, the next step is not ``end a QA session'' but instead goes back to ``sort remaining''.
    ]{}
    \label{fig:single-edit-workflow}
\end{figure*}

\subsection{Datasets to be Quality-Assured}
\label{sec:QA-datasets}

In the simulation, we consider three real datasets to be quality-assured from the \QAforML application scenarios introduced in Section~\ref{sec:application}.

\textbf{Visual Object Detection:}
This scenario concerns detecting visual objects from historical visualizations as introduced in Zhang et al.'s work~\cite{Zhang2021MI3} (see Figure~\ref{fig:application-workflows}($\mathrm{a_1}$)).
An ML model has assigned default labels to 3975 candidates of visual objects in John Snow's cholera map~\cite{Snow1855Mode}.
A post hoc analysis reveals that the model predictions have a normalized confusion matrix
$cm = \big(\begin{smallmatrix}
            86.98\% & 0.87\%\\
            0.53\% & 11.61\%
        \end{smallmatrix}\big)$,
i.e., $\frac{n_{TN}}{n} = 86.98\%$, $\frac{n_{FP}}{n} = 0.87\%$, $\frac{n_{FN}}{n} = 0.53\%$, and $\frac{n_{TP}}{n} = 11.61\%$.
We refer to this dataset as \emph{JS-block}.

\textbf{Visual Object Grouping:}
This scenario concerns classifying pairs of visual objects to group visual objects as introduced in Zhang et al.'s work~\cite{Zhang2021MI3} (see Figure~\ref{fig:application-workflows}($\mathrm{a_2}$)).
An ML model has assigned default labels to 782 candidates of visual object pairs to determine whether they should be grouped.
A post hoc analysis reveals that the model predictions have a normalized confusion matrix
$cm = \big(\begin{smallmatrix}
            70.15\% & 0.12\%\\
            0.36\% & 29.37\%
        \end{smallmatrix}\big)$.
We refer to this dataset as \emph{JS-grouping}.

\textbf{Solar Panel Detection:}
This scenario concerns classifying aerial images to detect solar panels (see Figure~\ref{fig:application-workflows}(b)).
This dataset is provided by Statistics Netherlands\footnote{\url{https://www.cbs.nl/}}.
An ML model assigned default labels to 1278 aerial images.
A post hoc analysis reveals that the model predictions have a normalized confusion matrix
$cm = \big(\begin{smallmatrix}
            64.82\% & 5.60\%\\
            5.44\% & 24.14\%
        \end{smallmatrix}\big)$.
We refer to this dataset as \emph{SP-image}.

\subsection{Conceptual Model for Simulation}
\label{sec:factor-layout-application-concept}

Figure~\ref{fig:single-edit-workflow} shows a \QAforML \workflow{}.
Consider a grid panel with $K$ grid cells.
$K_{FP}$ and $K_{FN}$ are the numbers of false-positive and false-negative labels shown in the grid panel.
The grid panel provides two commands for the user: \emph{request a new batch} and \emph{make a single edit}.
Assuming that the user operations are sequential, the total time cost of a round of quality assurance for the $K$ grid cells is a summation of the user operation time costs:

\begin{equation}\label{eq:basic-sum}
    T_{round} = t_{new} + K \cdot t_{view} + (K_{FP} + K_{FN}) \cdot t_{single}
\end{equation}

For a session with $n$ data objects, denote its overall confusion matrix as
$\big(\begin{smallmatrix}
            n_{TN} & n_{FP}\\
            n_{FN} & n_{TP}
        \end{smallmatrix}\big)$.
One can extrapolate $T_{round}$ and deduce the total time cost of a session:
$ T_{session} = \lceil \frac{n}{K} \rceil \cdot t_{new} + n \cdot t_{view} + (n_{FP} + n_{FN}) \cdot t_{single} $.
As long as we know the estimated values of operator time costs $t_{new}$, $t_{view}$, and $t_{single}$, we can estimate $T_{round}$ and $T_{session}$.

\subsection{Estimating Operator Time Costs}
\label{sec:factor-layout-application-estimation}

There are various approaches to estimating operator time costs $t_{new}$, $t_{view}$, and $t_{single}$.
One may recruit potential users to use a \QAforML interface and use eye-tracking, video recording, or interaction logs to capture detailed timing data.
However, such processes usually require a relatively complex experiment setup unavailable or inconvenient to most software development projects.
Another common practice in the model-based evaluation literature is to reuse time costs estimated by previous work~\cite{Card1980Keystroke}.
To accomplish a swift evaluation, one may also plug in hypothetical values for the operator time costs, such as assuming all types of operations cost the same.
While these approaches are all feasible for the \QAforML simulation, in the following, we introduce an approach to get a realistic estimation of operator time costs swiftly.

\subsubsection{Method}

We conduct self-experimentation under different conditions for $n_{batch}$, $K_{FP} + K_{FN}$, $layout$, and $application$ to gather observations of $T_{round}$ values.
Then, we use numerical analysis and modeling, such as multiple linear regression, to obtain $t_{new}$, $t_{view}$, and $t_{single}$ values.
The UI specialists can use the obtained models to inform conditions (e.g., layout designs) that have not been experimented with.
Below, we outline the process to model $t_{new}$, $t_{view}$, and $t_{single}$.
More details can be found in
\ifx\hideappendix\undefined
    Appendix~\ref{app:timing-factor}.
\else
    Appendix A in the supplementary materials.
\fi

\begin{figure}[!htbp]
    \centering
    \includegraphics[width=\linewidth]{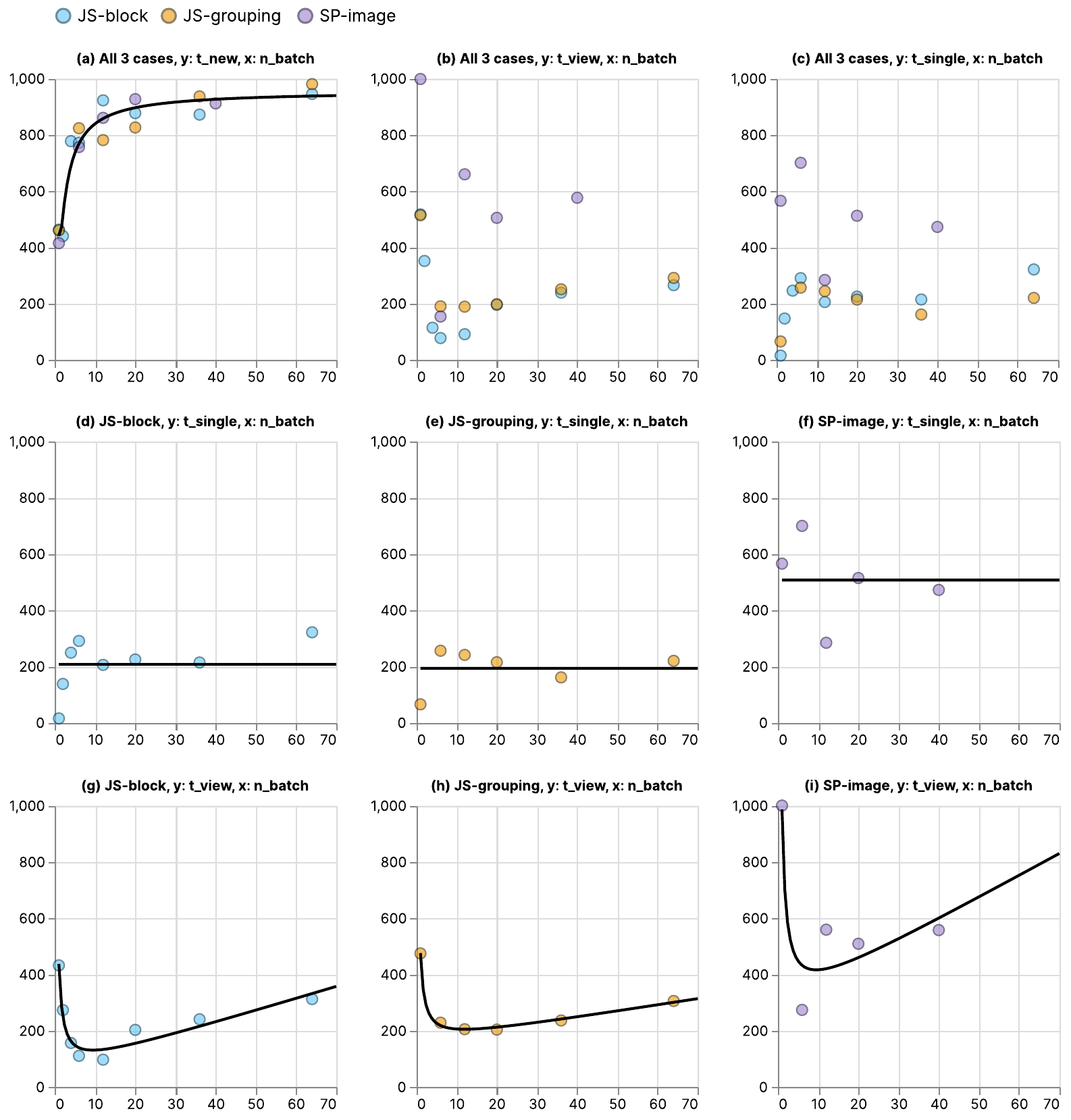}
    \caption{
        Summary of the experiment result and fitted models:
        (a) Estimations of $t_{new}$ by multiple linear regression.
        The selected model of $t_{new}$ that we plot is $ y = a + b/x + c/x^2 $.
        (b) Initial estimations of $t_{view}$ by multiple linear regression.
        (c) Initial estimations of $t_{single}$ by multiple linear regression.
        (d - f) Models of $t_{single}$ for the three applications.
        The selected model is $ y = a $.
        The data points are $t_{single}$ reestimated by putting $t_{new}$ back.
        (g - i) Models of $t_{view}$ for the three applications.
        The selected model is $ y = a + bx + c/x $. The data points are $t_{view}$ reestimated by putting $t_{new}$ and $t_{single}$ back.
        The x-axes of the subfigures are the number of grid cells $n_{batch}$.
        The y-axes of the subfigures are the corresponding operator time costs $t_{new}$, $t_{view}$, or $t_{single}$.
        The unit of all the time costs is milliseconds.
    }
    \Description[
        This figure shows the experiment result and fitted models.
        The figure has three rows and three columns with nine subfigures.
        Each subfigure shows a scatterplot with the horizontal axis denoting the number of grid cells in the experimented interface and the vertical axis denoting the operator time costs.
        The first row contains three subfigures, (a), (b), and (c).
        They show the distributions of all the data points for t new, t view, and t single, respectively.
        For t new, the three applications, JS-block, JS-grouping, and SP-image, exhibit a similar trend of first increasing drastically and then increasing slowly.
        The second row contains three subfigures (d), (e), and (f).
        They show the distributions of the t single data points for the three applications, JS-block, JS-grouping, and SP-image.
        The three applications' data points are distributed close to a horizontal line.
        The third row contains three subfigures (g), (h), and (i).
        They show the distributions of the t view data points for the three applications, JS-block, JS-grouping, and SP-image, respectively.
        The data points are distributed following U-shaped curves for the three applications.
    ]{}
    \label{fig:operator-time-costs}
\end{figure}

\subsubsection{Procedure and Results}

The values of operator time costs $t_{new}$, $t_{view}$, and $t_{single}$ may depend on the \emph{layout} of the \QAforML interface and the \emph{application}.
For example, the time to view each grid cell, i.e., $t_{view}$, may vary for different sizes of grid cells in the editing panel (dependent on layout), and the difficulty of recognizing the data objects shown in these grid cells (dependent on application).
Therefore, we estimate operator time costs separately for each layout and application.

\begin{figure}[!htbp]
    \centering
    \includegraphics[width=\linewidth]{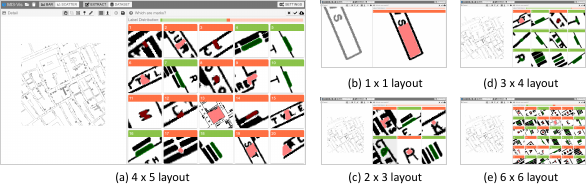}
    \caption{
        Different layouts of a grid-based interface for visual object detection in the JS-block dataset:
        (a) $4 \times 5$ layout.
        (b) $1 \times 1$ layout.
        (c) $2 \times 3$ layout.
        (d) $3 \times 4$ layout.
        (e) $6 \times 6$ layout.
    }
    \Description[
        The images show different layouts of the QA interface for the JS-block dataset.
    ]{}
    \label{fig:interface-layouts}
\end{figure}

For each application scenario introduced in Section~\ref{sec:QA-datasets}, we create variations of \QAforML interfaces with different layouts.
Figure~\ref{fig:interface-layouts} shows a few layout options of the \QAforML interface for the JS-block dataset.
We have experimented with eight grid layouts for JS-block ($1 \times 1$, $1 \times 2$, $2 \times 2$, $2 \times 3$, $3 \times 4$, $4 \times 5$, $6 \times 6$, and $8 \times 8$), six layouts for JS-grouping ($1 \times 1$, $2 \times 3$, $3 \times 4$, $4 \times 5$, $6 \times 6$, and $8 \times 8$), and five layouts for SP-image ($1 \times 1$, $2 \times 3$, $3 \times 4$, $4 \times 5$, and $5 \times 8$).
For each grid layout, we take a minimum of seven samples of $T_{round}$ with different values for $K$ and $K_{FP} + K_{FN}$
\ifx\hideappendix\undefined
    (see Table~\ref{tab:experiment-design-a} and Table~\ref{tab:experiment-design-b}).
\else
    (see Table 1 and Table 2 in Appendix A).
\fi

We assume $T_{round}$ follows Equation~\ref{eq:basic-sum}.
Thus, for each combination of layout and application, given the observations with different $K$ and $K_{FP} + K_{FN}$, we use multiple linear regression to estimate $t_{new}$, $t_{view}$, and $t_{single}$ values.
The estimation results are shown in Figure~\ref{fig:operator-time-costs}(a - c) and numerically in
\ifx\hideappendix\undefined
    Table~\ref{tab:experiment-summary}.
\else
    Table 3 in Appendix B.
\fi
The estimations of the user operation time costs in
\ifx\hideappendix\undefined
    Table~\ref{tab:experiment-summary}
\else
    Table 3 in Appendix B
\fi
are mostly statistically significant with high $R^2$, which suggests that our linear model in Equation~\ref{eq:basic-sum} is numerically accurate.
Thus, our assumption that the user operations are sequential instead of parallel is suitable for modeling.

Then, for each application, we fit a function of the operator time costs regarding layout (parameterized as $n_{batch}$).
Our purpose in fitting the function is not to obtain an ultimate model of the time costs.
Using the fitted functions, we can obtain smoothed estimations of the operator time costs and predict the operator time costs for the layouts we have not experimented with.

The $t_{new}$ values for the three application scenarios are shown in Figure~\ref{fig:operator-time-costs}(a) with three colors.
They show similar trends concerning the number of grid cells.
Based on this observation and the rationale that the time required to issue a new batch command is likely similar among the three application scenarios, we have decided to fit a single model to the three sets of data points.

\subsubsection{Details on Curve Fitting}

Firstly, we use a set of elementary functions (e.g., constant, polynomial, exponential, logarithm) to fit the operator time costs as a function of $n_{batch}$.
Then, we make attempts to formulate a semantic explanation of those functions with low error measures and abandon the functions for which we cannot find an intuitive explanation.
We choose a function as the final model based on fitting errors and interpretability.
Fitting the functions enables us to smooth the estimations of the operator time costs.
In the following, we introduce the process of curve fitting.
More details can be found in
\ifx\hideappendix\undefined
    Appendix~\ref{app:timing-factor} and Appendix~\ref{app:reestimation}.
\else
    Appendix A and Appendix B.
\fi

We use $n_{batch}$ to parameterize the layout.
One may expect that for different layouts (e.g., $2 \times 3$ and $3 \times 2$) with the same $n_{batch}$, the operator time costs may still be different.
Thus, it is possible to fit functions with two parameters, the number of rows and columns.
We use only one parameter, $n_{batch}$, because the ratio of the number of rows and number of columns in all the layouts we have experimented with are close and are all between $1 : 1$ and $1 : 2$.

\textbf{Fitting $t_{new}$:}
For the initial estimations of $t_{new}$ in Figure~\ref{fig:operator-time-costs}(a), formulae $a + b/x$, $a + b/x + c/{x^2}$, and $a + b/x + c/{x^2} + d/{x^3}$ are found by numerical analysis to be among the best candidate models for two-, three-, and four-parameter models, respectively.
With a positive $a$ and a negative $b$, these formulae suggest that there may be a maximum time cost in activating a new batch.
A smaller number of grid cells can reduce the cost, possibly because of the easiness of deciding if a QA round is completed.
The curve shown in Figure~\ref{fig:operator-time-costs}(a) is
$ \mathbf{t_{new} = 958.0085 - \frac{1254.8737}{n_{batch}} + \frac{739.4750}{n_{batch}^2}} $

As shown in Figure~\ref{fig:operator-time-costs}(b), the $t_{view}$ values estimated based on Equation~\ref{eq:basic-sum} exhibit different patterns for the three application scenarios.
This is understandable, as viewing and comprehending the three types of data objects may incur different time costs.
In particular, viewing solar panel images is generally more time-consuming, except that the second data point seems to be an outlier.
As shown in Figure~\ref{fig:operator-time-costs}(c), the estimated $t_{single}$ values for JS-block and JS-grouping are similar.
In contrast, the $t_{single}$ values for SP-image are generally higher than those of JS-block and JS-grouping.

We thus separately model $t_{view}$ and $t_{single}$ for the three application scenarios.
We choose to model $t_{single}$ next because we expect $t_{single}$ to be less affected by the size of the data objects being displayed than $t_{view}$.
The time cost to issue a single edit command, $t_{single}$, is mainly about the user's human motor control.
The time cost to view a data object, $t_{view}$, requires the user's comprehension.
Moreover, there are fewer single edit actions than viewing actions, and modeling $t_{single}$ is expected to affect $t_{view}$ less than modeling $t_{view}$ before $t_{single}$.

\textbf{Fitting $t_{single}$:}
Before modeling $t_{single}$, we reestimated the $t_{single}$ data points by replacing the $t_{new}$ variable in Equation~\ref{eq:basic-sum} with modeled $t_{new}$ values.
The reestimated $t_{single}$ data points for the three application scenarios are shown in Figure~\ref{fig:operator-time-costs}(d - f) respectively.
Following the same procedure to gather and analyze candidate models, we have selected a constant model for each application scenario.
The three models are:

\begin{itemize}
    \item \textbf{For JS-block:} $ \mathbf{t_{single} = 208.0818} $
    \item \textbf{For JS-grouping:} $ \mathbf{t_{single} = 193.8312} $
    \item \textbf{For SP-image:} $ \mathbf{t_{single} = 507.9517} $
\end{itemize}

\textbf{Fitting $t_{view}$:}
In the third iteration, we first reestimate the $t_{view}$ data points using the modeled $t_{new}$ and $t_{single}$ data.
As shown in Figure~\ref{fig:operator-time-costs}(g - i), the modeling of $t_{new}$ and $t_{single}$ hardly caused a noticeable displacement of the $t_{view}$ data points.
We can also easily observe a common phenomenon that the $t_{view}$ values are higher when there are too many or too few grid cells (equivalently, when the sizes of grid cells are too small or too large).
Following the same procedure for gathering and analyzing candidate models, we select the formula $y=a+bx+c/x$ for all three application scenarios as it semantically captures the effect of grid sizes with its second and third terms.
The three models are:

\begin{itemize}
    \item \textbf{For JS-block:} $ \mathbf{t_{view} = 50.4283 + 4.3202 n_{batch} + \frac{383.5033}{n_{batch}}} $
    \item \textbf{For JS-grouping:} $ \mathbf{t_{view} = 151.9951 + 2.2618 n_{batch} + \frac{322.6178}{n_{batch}}} $
    \item \textbf{For SP-image:} $ \mathbf{t_{view} = 267.0109 + 7.9103 n_{batch} + \frac{712.8504}{n_{batch}}} $
\end{itemize}

Following the curve fitting and reestimation steps, we obtain the operator time costs in
\ifx\hideappendix\undefined
    Table~\ref{tab:experiment-summary-reestimate}.
\else
    Table 4 in Appendix B.
\fi

\subsubsection{Findings}

Through the operator time cost estimations, we have the following observations:

\begin{itemize}
    \item The initial estimations of the operator time costs exhibit low square errors as shown in
          \ifx\hideappendix\undefined
              Table~\ref{tab:experiment-summary}.
          \else
              Table 3 in Appendix B.
          \fi
          The estimation is conducted with multiple linear regression because we assume the user operations are sequential and the total time cost is a summation of the operator time costs.
          The low square errors imply that the assumption is reliable, at least for numeric modeling purposes.

    \item The operator time costs depend on layout and application, as shown in Figure~\ref{fig:operator-time-costs}.
          It implies that \QAforML interface designers need to optimize the layout to improve users' productivity.

    \item $\mathbf{t_{view}}$ follows a U-shaped trend with the number of grid cells in the grid layout (as shown in Figure~\ref{fig:operator-time-costs}(g - i)).
          We interpreted the trend as grid cells that are too large or too small are hard to view.
          The user may need to inspect small grid cells more carefully, which costs more time.
          For large grid cells, we interpret that the user may need additional pupil movements for the foveal vision to cover the visual representation of the data object.

    \item The estimations of $\mathbf{t_{view}}$ in SP-image are much larger than those in JS-block and JS-grouping.
          We conceive it is harder for the user to comprehend the aerial images in SP-image than the visual objects in JS-block and JS-grouping.

\end{itemize}

\subsection{Simulations and Observations}
\label{sec:factor-layout-application-simulation}

\begin{algorithm}[htbp]
    \small
    \caption{Grid-based Interface (Single Edit)}\label{alg:grid-based-interface-single-edit}
    \begin{algorithmic}[1] %
        \Require true labels $ Y^* $, default labels $ Y $
        \Ensure user operation sequence $ ops = (op_1, ..., op_m) $
        \State $ \mathrm{i \gets 1} $
        \State $ \mathrm{ops \gets []} $
        \Do
            \State ops.push(``new'') \Comment{request a new batch}
            \State $ \mathrm{batchEnd \gets min(i + n_{batch}, n)} $
            \For{$ \mathrm{j \in (i, i+1, ..., batchEnd)} $}
                \State ops.push(``view'') \Comment{view an object}
                \If{$ \mathrm{Y[j] \ne Y^*[j]} $} \Comment{mislabeled?}
                    \State ops.push(``single'') \Comment{make a single edit}
                \EndIf
            \EndFor
            \State $ \mathrm{i = i + n_{batch}} $
        \DoWhile{$ \mathrm{i \leq n} $} \Comment{session ends?}
        \State \Return{$ \mathrm{ops} $}
    \end{algorithmic}
\end{algorithm}

In the following, we use simulation to examine the influence of application and layout on the total time cost $T_{session}$.
We demonstrate using simulation to find the optimal layout of \QAforML interfaces.

\subsubsection{Simulation Setup}

We use simulations to estimate $T_{session}$ for different applications and layouts following the \workflow{} in Figure~\ref{fig:single-edit-workflow}.
Algorithm~\ref{alg:grid-based-interface-single-edit} describes how to simulate an operation sequence in a simulation trial.
We run the simulations with the following parameter values:

\begin{itemize}
    \item \textbf{Interface layout:} all the $layout$ for which we have estimated operator time costs in Section~\ref{sec:factor-layout-application-estimation}.

    \item \textbf{Dataset:} three conditions corresponding to the dataset size $n$ and normalized confusion matrix $cm$ of the three applications JS-block, JS-grouping, and SP-image introduced in Section~\ref{sec:QA-datasets}.

    \item \textbf{Operator time costs:} $t_{new}$, $t_{view}$, and $t_{single}$ to be the values measured for the given $application$ and $layout$ as shown in Figure~\ref{fig:operator-time-costs} and
          \ifx\hideappendix\undefined
              Table~\ref{tab:experiment-summary-reestimate}.
          \else
              Table 4 in Appendix B.
          \fi
\end{itemize}

\begin{figure}[!htbp]
    \centering
    \includegraphics[width=\linewidth]{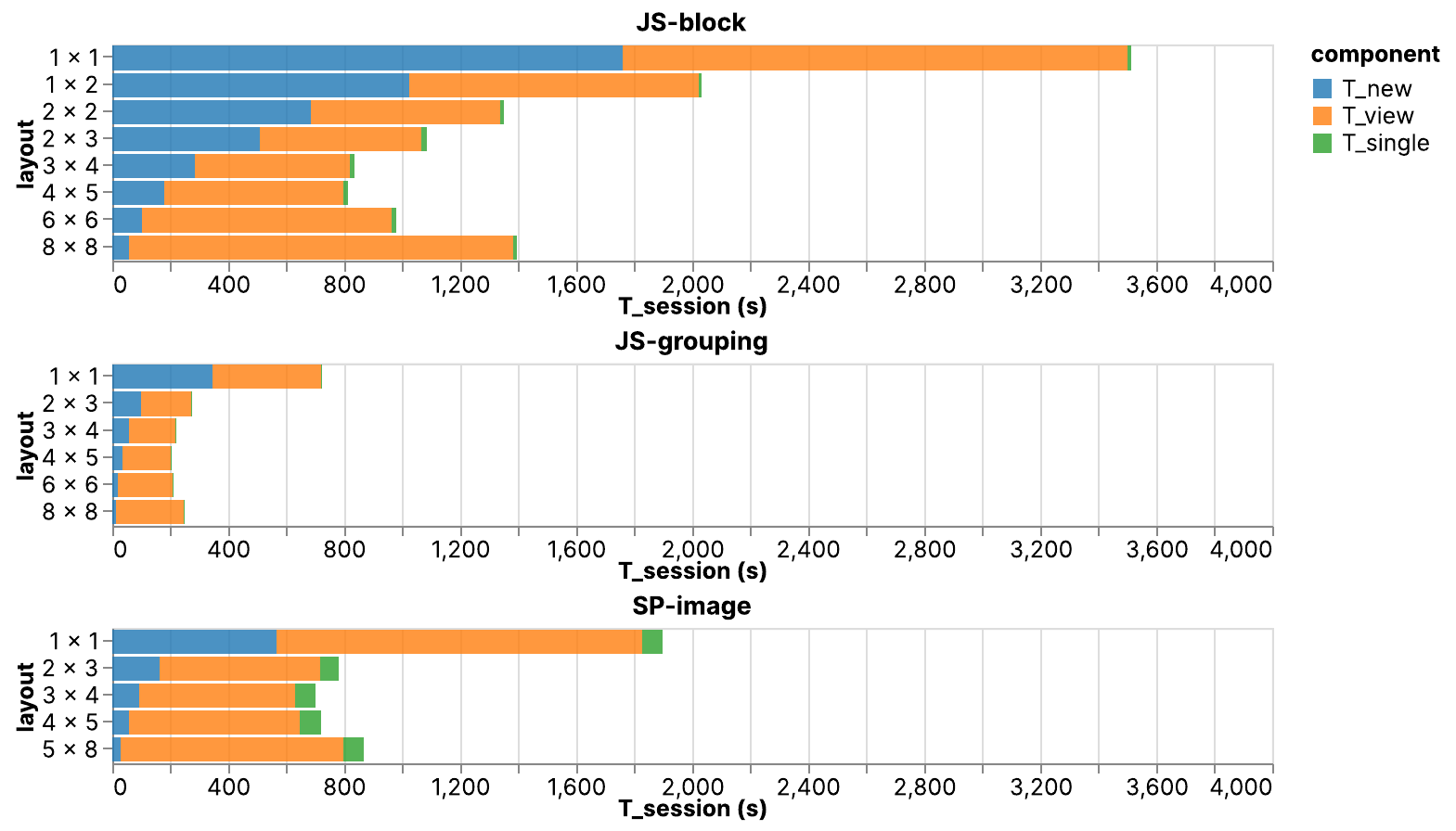}
    \caption{
        Simulation of layouts:
        The effect of grid layout in the three applications.
        For each application, the total time cost follows a U-shaped curve.
        For JS-block and JS-grouping, the minimal is achieved at $ 4 \times 5 $.
        For SP-image, the minimal is achieved at $ 3 \times 4 $.
    }
    \Description[
        Fully described in the caption.
    ]{}
    \label{fig:simulation-layout}
\end{figure}

\subsubsection{Result and Analysis}

Figure~\ref{fig:simulation-layout} shows the simulation results in relation to different grid layouts and applications.
The bar charts in the three applications all exhibit a U-shape pattern.
\emph{For JS-block and JS-grouping, the minimal total time cost for a QA session is achieved with the $ 4 \times 5 $ layout.}
\emph{For SP-image, the minimum is achieved with the $ 3 \times 4 $ layout.}

The default labels have low error rates (1.4\% for JS-block, 0.48\% for JS-grouping, and 11.04\% for SP-image).
Thus, the total cost of single edit operations ($T_{single}$) is low compared with the total cost of the operations for starting a new batch ($T_{new}$) and the total cost of the viewing actions ($T_{view}$).
The optimal layout reflects a trade-off between $T_{new}$ and $T_{view}$.

Layout significantly impacts $T_{session}$, as shown in Figure~\ref{fig:simulation-layout}.
The optimal layout depends on the QA application.
\emph{No single layout is optimal for all the application scenarios.}
A design implication is that for \QAforML interfaces, UI specialists may need to optimize the layout separately for each application.
Alternatively, \QAforML interfaces may provide the user with layout-switching functions.

  \section{The Factor of User Label Strategy}
\label{sec:factor-label-strategy}

Section~\ref{sec:factor-layout-application} has introduced the setting where using ``single edit'' commands is the only way to edit labels.
For the grid-based interfaces, except for the $ 1 \times 1 $ layout, all the other layouts present multiple data objects simultaneously.
Thus, a natural extension of the interface functionality is to provide ``batch edit'' commands that simultaneously edit the label of multiple data objects.
Users may use their intelligence to take shortcuts when batch edit commands are available.

This section uses simulation to examine the benefit of providing batch edit commands.
When batch edit commands are available, the user has multiple methods to accomplish a QA session, such as using only the single edit commands (as is in Section~\ref{sec:factor-layout-application}) or combining single and batch edit commands.
Thus, we model the user's label strategy to choose between the methods.

The JS-block and JS-grouping applications concern active learning \workflow{}s, as illustrated in Figure~\ref{fig:application-workflows}(a).
A user can dynamically decide when to stop the active learning after labeling $k$ data objects interactively.
If $k$ is small, the cost of manual labeling during active learning is low, but the ML model is likely less accurate, and more erroneous labels are to be corrected during \QAforML.
On the other hand, the less accurate ML results are, the more benefit the batch edit commands may bring.
Hence, it is interesting to simulate the impact of the user's strategies for using batch edit commands in conjunction with different accuracy levels of the ML results.

\subsection{Conceptual Model for Simulation}
\label{sec:factor-label-strategy-concept}

After introducing the batch edit commands, the user has multiple methods to carry out the QA processes.
Moreover, the user may not use the same method for all the batches.
Instead, the user may decide to apply different methods depending on the distribution of TP, TN, FP, and TN in each batch.
In this section, we model the user's strategy of choosing among the methods.
The user strategy is essentially the ``selection rules'' in GOMS models~\cite{John1996GOMS}.

\subsubsection{Batch Edit Methods}

Figure~\ref{fig:batch-edit-workflow} extends the basic \QAforML workflow in Figure~\ref{fig:single-edit-workflow} with batch editing commands.
For the quality assurance of binary classifications, we consider three batch edit commands:

\begin{itemize}
    \item \textbf{All positive:} label the data objects ``positive''.
    \item \textbf{All negative:} label the data objects ``negative''.
    \item \textbf{Inverse all:} label the data objects to be the logical inverse of their current labels.
\end{itemize}

When quality-assuring the labels of a batch of data objects, the batch edit commands should be used at most once if a user makes no mistakes in observation and command selection.
When the three batch edit commands are available, a user can observe all data objects in the editing panel and choose one of the following four methods to finish a round of quality assurance.
Each method has a different time cost $T_{round}$.

\begin{itemize}
    \item \textbf{Baseline (B):} using only ``single edit'' commands to correct all the false-positive/negative errors (number of single edits $N_{single} = K_{FP} + K_{FN}$).
          In this case, $ T_{round} = T_0 + (K_{FP} + K_{FN}) \cdot t_{single} $.
    \item \textbf{Mostly positive (P):} issuing an ``all positive'' command, and then using ``single edit'' commands to correct the remaining errors ($N_{single} = K_{FP} + K_{TN}$).
          In this case, $ T_{round} = T_0 + t_{allPositive} + (K_{FP} + K_{TN}) \cdot t_{single} $.
    \item \textbf{Mostly negative (N):} issuing an ``all negative'' command, and then using ``single edit'' commands to correct the remaining errors ($N_{single} = K_{FN} + K_{TP}$).
          In this case, $ T_{round} = T_0 + t_{allNegative} + (K_{FN} + K_{TP}) \cdot t_{single} $.
    \item \textbf{Mostly wrong (W):} issuing an ``inverse all'' command, and then using ``single edit'' commands to correct the remaining errors ($N_{single} = K_{TP} + K_{TN}$).
          In this case, $ T_{round} = T_0 + t_{inverseAll} + (K_{TP} + K_{TN}) \cdot t_{single} $.
\end{itemize}

In the time costs, $T_0 = t_{new} + t_{overview} + K \cdot t_{view}$.
The operator time costs, $t_{overview}$, $t_{allPositive}$, $t_{allNegative}$, and $t_{inverseAll}$, are the time costs to overview all grid cells in the editing area and issue one of the three batch edit commands, respectively.
Which of the four methods is the most efficient (with the minimal $T_{round}$) depends on the confusion matrix
$\big(\begin{smallmatrix}
            K_{TN} & K_{FP}\\
            K_{FN} & K_{TP}
        \end{smallmatrix}\big)$
of the batch of data objects.

The set of possible labeling methods depends on the available commands (denoted as $cmds$) for editing labels.
Typically, ``single edit'' commands are available to ensure that each label can be edited individually.
Thus, we assume $SingleEdit \in cmds$.
When a batch edit command ( ``all positive'', ``all negative'', or ``inverse all'') is unavailable in the \QAforML interface, the corresponding labeling method (P, N, or W) is unavailable.
For a \QAforML interface without batch edit commands, we refer to its $cmds$ as $cmds_{single} = \{SingleEdit\}$.
When all four label edit commands are available, we refer to its $cmds$ as $cmds_{batch} = \{SingleEdit, AllPositive, AllNegative, InverseAll\}$.

\subsubsection{Parameterizing User Label Strategy}

The effectiveness of the label edit commands depends on a user's ability to choose the right method that minimizes the time cost when facing a batch of data objects.
We parameterize this factor of a user's ability to make strategic choices so that we can simulate different levels of the user's ability.
The parameterization of user strategies corresponds to an implementation of ``SelectEditCommand'' in Algorithm~\ref{alg:session-simulation}.
We introduce two model parameters:

\begin{itemize}
    \item \textbf{User skill level:} $u_{sl} \in [-1, 1]$ represents different levels of user skills, with 1 being the best and -1 being the worst.
    \item \textbf{User strategy uncertainty:} $u_{su} \in [0, 1]$ represents randomness in choosing among the methods, with 1 being entirely random and 0 being deterministic.
\end{itemize}

The overall idea of the parameterization is that the labeling methods can be ranked by time costs.
A higher skill level corresponds to choosing a smaller rank index.
The choice of a rank index may be associated with an error range.
The strategy uncertainty captures the width of the error range.

When $u_{sl} = 0$, a user does not use batch edit commands and always chooses the baseline method B.
When $u_{sl} = 1$, the user is highly experienced and always chooses the best method for each batch.
Let $N_m $ be the total number of methods.
$N_m = 4$ when the batch edit commands are all available.
$N_m = 1$ when only single edit commands are available.
A simulation model knows the ground truth labels.
Therefore, it can sort all labeling methods by time costs from 0 to $N_m - 1$, with the 0-th being the best.
The strategy of always choosing the 0-th method corresponds to $u_{sl} = 1$.
The strategy of always choosing the $(N_m - 1)$-th method corresponds to $u_{sl} = -1$.

When $0 < u_{sl} < 1$, the simulation model chooses the $k$-th labeling method between the best method at position 0 and the baseline method at position $bl$, such that $k = \mathrm{round} \big( (1-u_{sl}) \cdot bl \big)$.
If the baseline method is the best, $bl = 0$ and $k = 0$.

When $-1 < u_{sl} < 0$, the simulation model chooses the $k$-th labeling method between the baseline method at position $bl$ and the worst method at position $(N_m-1)$, such that $k = \mathrm{round} \big(- u_{sl} \cdot (N_m-1-bl)+bl \big)$.
Usually, users will not deliberately choose a batch edit command that would lead to more user operations than the baseline method.
It is thus rare to have $u_{sl} < 0$.

When $u_{su} = 1$, the selection is entirely random among all $N_m$ strategies.
When $u_{su} = 0$, the selection of the labeling method is deterministic and is based on $u_{sl}$ only.
When $0 < u_{su} < 1$, the simulation model randomly chooses a labeling method ranked between the $a$-th and $b$-th positions.
$a = \max(0,\;k-\delta)$, $b=\min(N_m-1,\;k+\delta)$, and $\delta = \mathrm{round}(u_{su} \cdot N_m)$.

\subsection{Estimating Operator Time Costs}
\label{sec:factor-label-strategy-estimation}

In principle, $t_{overview}$, $t_{allPositive}$, $t_{allNegative}$, and $t_{inverseAll}$ can be estimated in a manner similar to that for $t_{new}$, $t_{view}$, and $t_{single}$ in Section~\ref{sec:factor-layout-application}.
In the following, we consider a discount process for estimating these operator time costs.

It is intuitive to anticipate that $t_{allPositive}$, $t_{allNegative}$, and $t_{inverseAll}$ are close to $t_{new}$ because their corresponding buttons are all located at the header toolbar.
We thus assume $t_{allPositive} = t_{allNegative} = t_{inverseAll} = t_{new}$.
The frequency of using the commands ``all positive'', ``all negative'', and ``inverse all'' is much lower than other operations.
Thus, even if the estimations of $t_{allPositive}$, $t_{allNegative}$, and $t_{inverseAll}$ have minor imprecisions, the imprecisions have a limited impact on the estimation of the total time cost.

The time that a user spends for overviewing all data objects ($t_{overview}$) before issuing (or skipping) a batch edit command can be costly when $n_{batch}$ is large.
It likely increases with $n_{batch}$ as the user needs to comprehend more data objects.
After the overview, there may be a small reduction in the time used for examining each individual data object ($t_{view}$).
To capture these intuitions, in the following simulations, we assume a simple linear model $t_{overview} = n_{batch} \cdot 0.025$ with the unit being second to ensure $t_{overview}$ is monotonous with $n_{batch}$.
We discount $t_{view}$ by 0.0125 seconds when an overview is conducted.

\subsection{Simulations and Observations}
\label{sec:factor-label-strategy-simulations}

Through this set of simulations, we examine the influence of introducing batch edit commands and user's label strategies to $T_{session}$.

\subsubsection{Simulation Setup}

We run the simulations with the following parameter values:

\begin{itemize}
    \item \textbf{Interface layout}: 7 conditions of $layout$ being $\{ 1 \times 2, 2 \times 2, 2 \times 3, 3 \times 4, 4 \times 5, 6 \times 6, 8 \times 8 \}$.

    \item \textbf{Dataset}: $n = 1000$ and
          $cm = \big( \begin{smallmatrix}
                      0.05 & 0.05\\
                      0.45 & 0.45
                  \end{smallmatrix} \big)$.

    \item \textbf{Operator time costs}: $t_{new}$, $t_{view}$, $t_{overview}$, $t_{single}$, and $t_{batch}$ to be the values measured for the JS-block application for any given $layout$.

    \item \textbf{Label strategy}: five conditions of $<cmds, u_{sl}, u_{su}>$ corresponding to different label strategies
          \begin{itemize}
              \item \textbf{NoBC}: never uses batch edit commands ($cmds = cmds_{single}$, $u_{sl} = 0$, $u_{su} = 0$).
              \item \textbf{OPT}: choose optimal combinations of batch and single edit commands ($cmds = cmds_{batch}$, $u_{sl} = 1$, $u_{su} = 0$).
              \item \textbf{Random}: choose a random combination of batch and single edit commands ($cmds = cmds_{batch}$, $u_{sl} = 1$, $u_{su} = 1$).
              \item \textbf{TOP2}: choose randomly between the top 2 combinations of batch and single edit commands ($cmds = cmds_{batch}$, $u_{sl} = 1$, $u_{su} = 0.25$).
              \item \textbf{TOP3}: choose randomly among the top 3 combinations of batch and single edit commands ($cmds = cmds_{batch}$, $u_{sl} = 1$, $u_{su} = 0.5$).
          \end{itemize}
\end{itemize}

Ten repeated simulation trials with different random seeds are run for each combination of simulation parameters.
The random seed determines the random permutation of data objects.

\begin{figure}[!htbp]
    \centering
    \includegraphics[width=\linewidth]{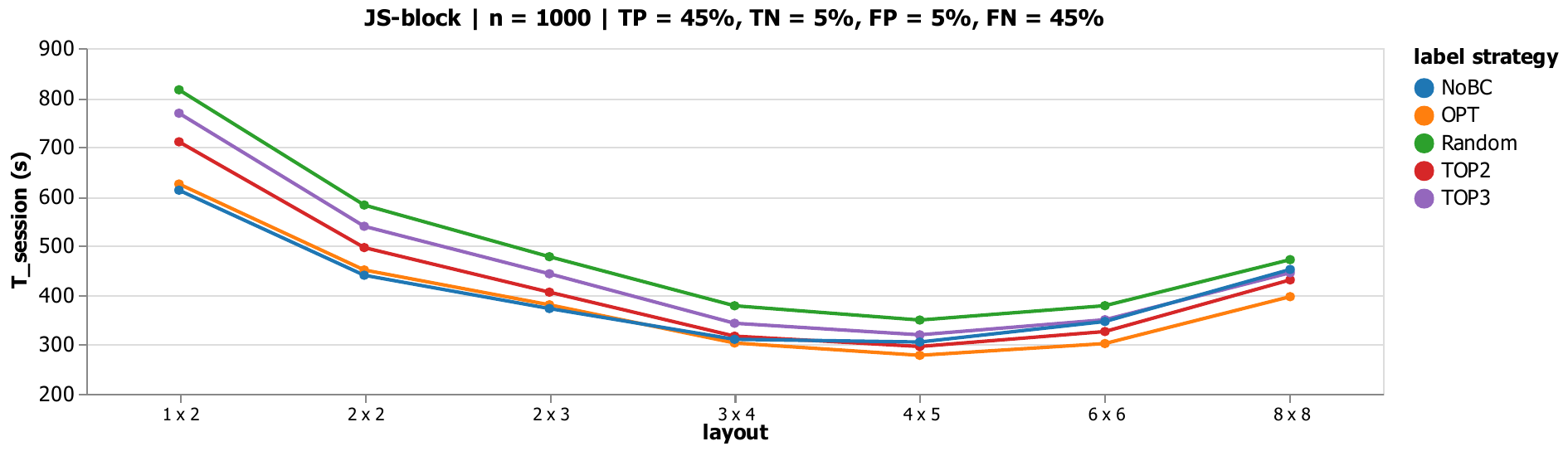}
    \caption{
        Simulation of label strategies:
        The simulation result of the total time cost $T_{session}$ with 5 different label strategies: NoBC, OPT, Random, TOP2, TOP3.
        The vertical axis starts from 200 instead of 0 to highlight the differences.
    }
    \Description[
        Fully described in the main text of the subsection.
    ]{}
    \label{fig:simulation-layout-label-strategy}
\end{figure}

\subsubsection{Result and Analysis}

As shown in Figure~\ref{fig:simulation-layout-label-strategy}, \emph{user's label strategy influences the time cost}.
For all the layouts with more grid cells than $2 \times 3$, OPT is consistently better than NoBC.
Note that NoBC can be better than OPT under some layouts.
The interface for NoBC has no batch edit commands, and thus NoBC can waive the time cost to overview the data objects, which is required for the other label strategies.
The minimal $T_{session}$ is achieved by OPT at $4 \times 5$ layout, which suggests that \emph{introducing batch edit commands may save user effort}.

Similar to Figure~\ref{fig:simulation-layout}, the time cost follows a U-shaped curve.
With the increase of $n_{batch}$, NoBC's relative performance consistently decreases compared with the other label strategies.
The advantage of using batch edit commands increases with $n_{batch}$.

  \section{The Factor of Default Label Accuracy}
\label{sec:factor-default-label-accuracy}

\QAforML \workflow{}s require the user to verify and correct machine predictions.
While it is intuitive that accurate machine predictions save user effort, it is not straightforward to what extent accurate predictions reduce task completion time.
This section investigates the impact of default label accuracy on the \QAforML time costs.
In practice, UI specialists may conduct such simulations to examine how much time can be saved by improving machine predictions.
In this way, UI specialists can decide whether it is worthwhile to improve the default label accuracy or whether it is more cost-effective to improve other aspects of the \QAforML interface.

\subsection{Simulations and Observations}

Through this set of simulations, we examine the influence of default label accuracy to $T_{session}$.

\subsubsection{Simulation Setup}

We run the simulations with the following parameter values:

\begin{itemize}
    \item \textbf{Interface layout:} five conditions of $layout$ being $\{ 1 \times 2, 2 \times 2, 2 \times 3, 3 \times 4, 4 \times 5, 5 \times 6, 8 \times 8 \}$.

    \item \textbf{Dataset:} $n = 1000$ and five conditions of $cm$ being
          $ \{ \big( \begin{smallmatrix}
                      0.5a & 0.5 - 0.5a\\
                      0.5 - 0.5a & 0.5a
                  \end{smallmatrix} \big) | a \in \{0.5, 0.6, 0.7, 0.8, 0.9\} \} $.
          The five conditions of $cm$ correspond to different default label accuracies: 0.5, 0.6, 0.7, 0.8, 0.9.

    \item \textbf{Operator time costs:} $t_{new}$, $t_{view}$, $t_{overview}$, $t_{single}$, and $t_{batch}$ to be the values measured for the JS-block application for any given $layout$.

    \item \textbf{Label strategy:} $cmds = cmds_{batch}$, $u_{sl} = 1$, and $u_{su} = 0$.
          With this configuration, the optimal label strategy is used for every batch.

\end{itemize}

Ten repeated simulation trials with different random seeds are run for each combination of simulation parameters.
The random seed determines the random permutation of data objects and the assignment of default labels.

\subsubsection{Result and Analysis}

\begin{figure}[htbp]
    \centering
    \includegraphics[width=\linewidth]{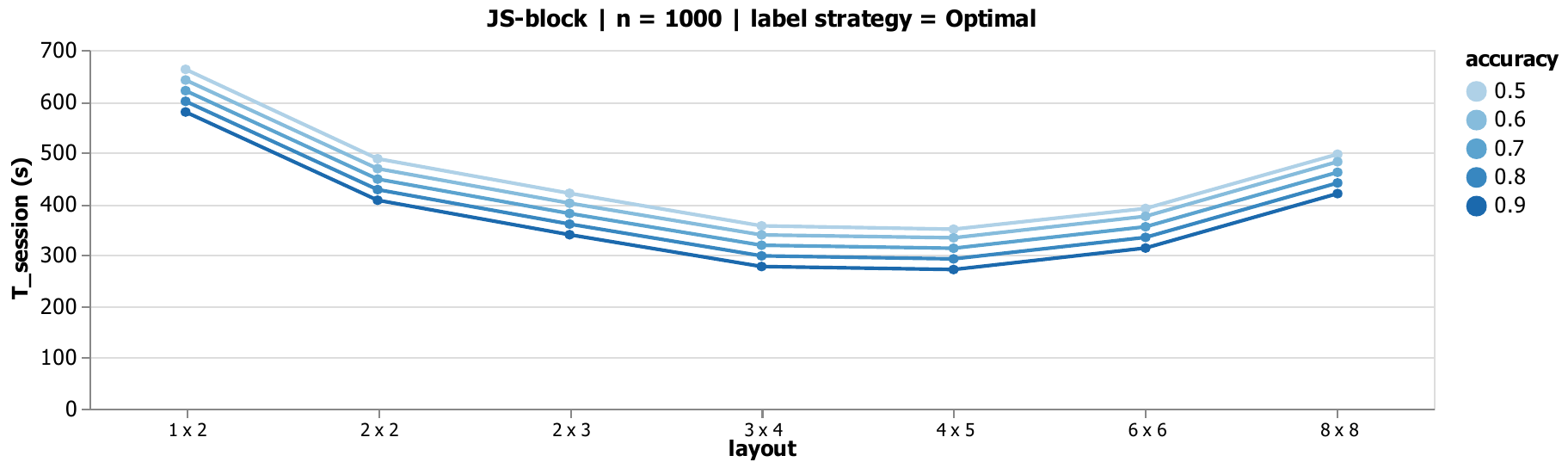}
    \caption{
        Simulation of default label accuracies:
        The simulation result of the total time cost $T_{session}$ with five conditions of default label accuracy (0.5, 0.6, 0.7, 0.8, and 0.9).
    }
    \Description[
        Fully described in the main text of the subsection.
    ]{}
    \label{fig:simulation-layout-default-label-accuracy}
\end{figure}

Figure~\ref{fig:simulation-layout-default-label-accuracy} shows the simulation result.
For all the layouts, \emph{with the increase of default label accuracy, the total time cost consistently decreases}.
Meanwhile, although accurate models save much time, using a good model (accuracy = 0.9) with a bad layout (1 $\times$ 2) may cost more time than using a bad model (accuracy = 0.5) with a good layout (3 $\times$ 4).
This observation implies that \emph{good interface and algorithm designs are both critical for saving user effort}.

  \section{The Factor of Rank Method}
\label{sec:factor-rank-method}

A general \QAforML \workflow{} involves rank methods to reorder data objects, as shown in Figure~\ref{fig:batch-edit-workflow}.
Rank methods can alter the local distributions of label categories and default labels' correctness in the batches.
Rank methods may reduce the time cost by unbalancing the local confusion matrix
$\big(\begin{smallmatrix}
            K_{TN} & K_{FP}\\
            K_{FN} & K_{TP}
        \end{smallmatrix}\big)$
of each batch.
The effectiveness of rank methods may affect the opportunities of using batch edit commands and influence the \QAforML time cost.

One may use actual implementations of the rank methods for simulation.
For example, Zhang et al.~\cite{Zhang2019Cost,Zhang2021MI3} simulate the number of user operations in active learning \workflow{}s.
Their simulations use the actual implementations of scoring functions in active learning to rank the data objects.
Meanwhile, using actual implementations can be inefficient.
The rank method needs to be implemented by the evaluator, which introduces an implementation cost.
Additionally, the actual implementation can be computationally expensive and slow down the simulation.

In the following, we introduce a parameterization of rank methods.
With the parameterization, we can simulate rank methods by configuring different parameter values.
The parameterization approach also gives us more control over the characteristics of the rank methods in simulations.

\subsection{Conceptual Model for Simulation}
\label{sec:factor-rank-method-concept}

To capture the rank methods in the simulation model, it is desirable to model their ability to change the local confusion matrix in each batch.
The modeling of the rank method corresponds to an implementation of ``Rank'' in Algorithm~\ref{alg:session-simulation}.
The following focuses on the rank methods that can be modeled as selecting a subset of data objects according to some criteria and moving them to the front of the list of data objects.
We refer to them as bipartition (``BiPart'') rank methods.

\subsubsection{Parameterizing Rank Method}

We parameterize a rank method with a matrix $ rm \in [0, 1]_{c \times c} $ where $c$ is the number of label categories.
$rm_{i, j}$ denotes the rate of data objects with true label $ i $ and default label $ j $ that are moved to the front of the list by the rank method among all such data objects.
The total number of data objects selected by the rank method $rm$ and moved to the front of the list is thus $ n_s = \sum_{i, j} rm_{i, j} \cdot n \cdot cm_{i, j} $.
All the selected data objects are regarded as unordered.
Thus, the selected data objects are randomly shuffled in the simulation.
For binary classification, let
$rm = \big(\begin{smallmatrix}
            p_{TN} & p_{FP}\\
            p_{FN} & p_{TP}
        \end{smallmatrix}\big)$
, we have $ n_s = n_{TP}p_{TP} + n_{TN}p_{TN} + n_{FP}p_{FP} + n_{FN}p_{FN} $.

\subsubsection{Examples}

Consider a rank method with
$rm = \big(\begin{smallmatrix}
            0 & 0\\
            1 & 1
        \end{smallmatrix}\big)$
that moves data objects with ground truth labels positive (i.e., TPs and FNs) to the front of the list.
For a batch containing a subset of these selected data objects, the user can use a batch edit command to set all their label categories to be ``Positive''.

Consider another rank method with
$rm = \big(\begin{smallmatrix}
            0 & 1\\
            1 & 0
        \end{smallmatrix}\big)$ that selects FNs and FPs.
Such a rank method moves all the data objects with incorrect default labels to the front, and all the user needs to do is to use a batch edit command to inverse all of their labels.

In these two examples, the rank methods are good as they create opportunities for the user to use the batch edit commands effectively.
A rank method can make the distributions of TP, TN, FP, and FN less uniform.
Any high concentration of one group of TP, TN, FP, or FN and combined groups of \{TP, FN\}, \{TN, FP\}, \{TP, TN\}, or \{FP, FN\} can be beneficial.
In these situations, batch edit commands are beneficial.
What is undesirable is any ordering result that groups a batch of data objects with similar numbers of TPs and FPs or similar numbers of TNs and FNs.

\subsubsection{Quantifying Rank Method Performance}

It is unrealistic to expect the rank method to be good enough to get all the TP/TN/FP/FN ranked top, as in the examples above.
We quantify the performance of a rank method by its difference from an ideal selection.
We denote an ideal selection with a binary matrix $ target \in \{0, 1\}_{c \times c} $.
In the matrix, $ target_{i, j} = 1 $ when data objects with true label $ i $ and default label $ j $ are intended to be selected.

A rank method $rm$ selects $n_s = \sum_{i, j} rm_{i, j} \cdot n \cdot cm_{i, j}$ data objects and moves them to the front of the list.
As specified by $ target $, among all the selected data objects, $ \sum_{i, j} rm_{i, j} \cdot n \cdot cm_{i, j} \cdot target_{i, j} $ are intended to be selected.
We use the rank method's precision, $ \frac{ \sum_{i, j} rm_{i, j} \cdot cm_{i, j} \cdot target_{i, j} }{ \sum_{i, j} rm_{i, j} \cdot cm_{i, j} } $, to quantify its performance.
Note that the precision depends on not only the rank method $rm$ itself but also the choice of the ideal selection $target$.

\subsection{Simulations and Observations}
\label{sec:factor-rank-method-precision-simulation}

Through this set of simulations, we examine the influence of introducing rank methods to $T_{session}$.

\subsubsection{Simulation Setup}

We run the simulations with the following parameter values:

\begin{itemize}
    \item \textbf{Interface layout:} five conditions of $layout$ being $\{ 2 \times 2, 2 \times 3, 3 \times 4, 4 \times 5, 6 \times 6, 8 \times 8 \}$.

    \item \textbf{Dataset:} $n = 1000$ and
          $cm = \big( \begin{smallmatrix}
                      0.50 & 0.05\\
                      0.30 & 0.15
                  \end{smallmatrix} \big)$.

    \item \textbf{Operator time costs:} $t_{new}$, $t_{view}$, $t_{overview}$, $t_{single}$, and $t_{batch}$ to be the values measured for the JS-block application for any given $layout$.

    \item \textbf{Label strategy:} $cmds = cmds_{batch}$, $u_{sl} = 1$, and $u_{su} = 0$.
          With this configuration, the optimal label strategy is used for every batch.

    \item \textbf{Rank method:} five conditions of $rm$ being
          $ \{ \big( \begin{smallmatrix}
                      \frac{1-a}{3} & \frac{1-a}{3}\\
                      a & \frac{1-a}{3}
                  \end{smallmatrix} \big)
              | a \in \{0, 0.25, 0.5, 0.75, 1\} \} $.
          We have also simulated a scenario where no rank method is used, and the dataset is randomly ordered.

\end{itemize}

We set
$target = \big(\begin{smallmatrix}
            0 & 0\\
            1 & 0
        \end{smallmatrix}\big)$,
i.e., the rank method's goal is to select false-negatives.
Thus, given
$cm = \big( \begin{smallmatrix}
            0.50 & 0.05\\
            0.30 & 0.15
        \end{smallmatrix} \big)$,
the BiPart rank method's precision is $\frac{9a}{2a + 7}$.
When $a$ equals 0, 0.25, 0.5, 0.75, and 1, the corresponding rank method precisions are (approximately) 0, 0.3, 0.56, 0.8, and 1.

Ten repeated simulation trials with different random seeds are run for each combination of simulation parameters.
The random seed determines the ordering of data objects and the assignment of default labels.

\begin{figure}[htbp]
    \centering
    \includegraphics[width=\linewidth]{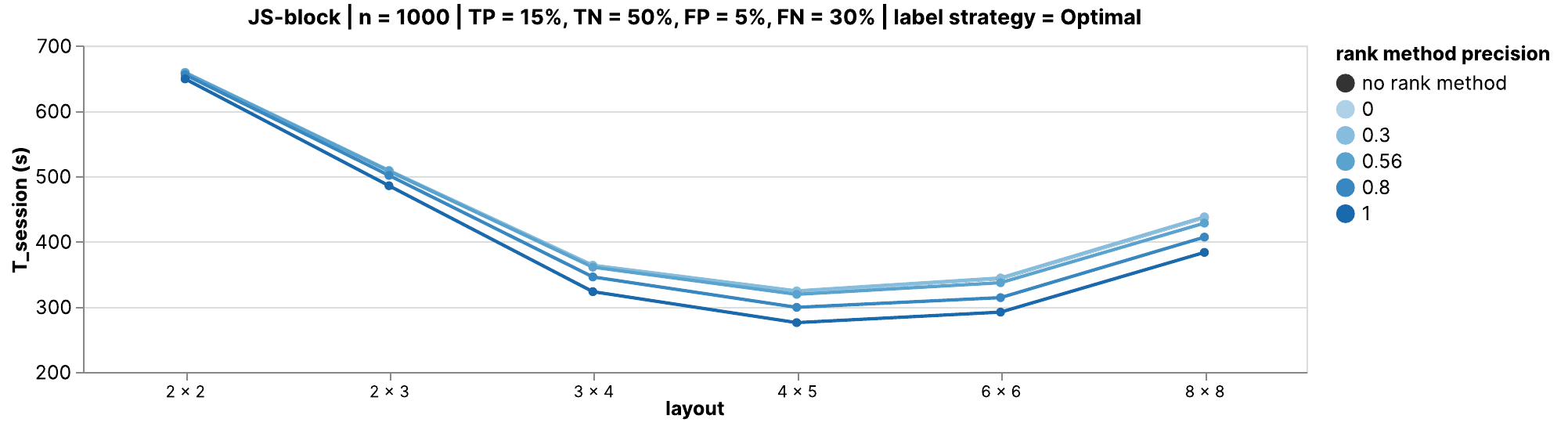}
    \caption{
        Simulation of rank method precisions:
        Five conditions of rank method precision are simulated, including 0, 0.3, 0.56, 0.8, and 1.
        An additional condition called ``no rank method'' is simulated where the dataset is randomly ordered.
        The curves of ``no rank method'' and rank methods with precision 0 and 0.3 heavily overlap.
        The vertical axis starts from 200 instead of 0 to highlight the differences.
    }
    \Description[
        Fully described in the main text of the subsection.
    ]{}
    \label{fig:simulation-layout-rank-method-precision}
\end{figure}

\subsubsection{Result and Analysis}

Figure~\ref{fig:simulation-layout-rank-method-precision} shows the simulation result.
Compared with not using rank methods, introducing the rank methods reduces the time cost.
For all the layouts, \emph{with the increase of rank method precision, the total time cost consistently decreases}.
Asides, the benefit of having rank methods with high precision is more prominent for layouts with larger $n_{batch}$.
For a larger $n_{batch}$, more data labels can be edited by activating one batch edit command.
The result provides the design implication that for \QAforML interfaces, \emph{it is beneficial to integrate rank methods with high precision to promote the use of batch edit commands}.

\section{Summary of the Simulation Model}
\label{sec:model-summary}

The simulation model described above is intended to be used by UI specialists.
The model can be used for different applications by specifying different simulation parameter values.
Table~\ref{tab:simulation-parameters} summarizes all the simulation parameters discussed in the previous sections.

\begin{table*}[htbp]
    \centering
    \footnotesize
    \caption{
        The parameters of our simulation model.
        We use standard math notations to denote the data types, such as $\mathbb{N}$ for integers, $\mathbb{R}$ for real numbers, + for $>0$, and $[a, b]$ for a range.
    }
    \begin{tabularx}{\linewidth}{lllX}
        \toprule
        \textbf{Parameter} & \textbf{Data Type}    & \textbf{Estimation} & \textbf{Description}                                                                                                                                          \\
        \midrule
        $n$                & $\mathbb{N}^+$        & external            & The number of data objects.                                                                                                                                   \\
        \hline
        $n_{batch}$        & $\mathbb{N}^+$        & observed            & The number of data objects of each batch.                                                                                                                     \\
        \hline
        $cm$               & $[0, 1]_{c \times c}$ & external            & The confusion matrix of default labels for the data to be quality-assured.                                                                                        \\
        \hline
        $t_{new}$          & $\mathbb{R}^+$        & captured            & The average time cost to activate the ``new batch'' command to fetch the next batch of data objects.                                                          \\
        \hline
        $t_{view}$         & $\mathbb{R}^+$        & captured            & The average time cost to view a data object and its label in a grid cell in the editing panel (including the cost of viewing the context panel if necessary). \\
        \hline
        $t_{overview}$     & $\mathbb{R}^+$        & captured            & The average time cost to overview all data objects in the editing panel (including the cost of viewing the context panel if necessary).                       \\
        \hline
        $t_{single}$       & $\mathbb{R}^+$        & captured            & The average time cost to activate a command to change the label of a data object in the editing panel.                                                        \\
        \hline
        $t_{batch}$        & $\mathbb{R}^+$        & captured            & The average time cost to activate a batch edit command to change the labels of all the data objects in the editing panel.                                     \\
        \hline
        $u_{sl}$           & $[-1, 1]$             & observed            & The user skill level in using batch edit commands.                                                                                                                \\
        \hline
        $u_{su}$           & $[0, 1]$              & observed            & The user strategy uncertainty in using batch edit commands.                                                                                                       \\
        \hline
        $cmds$             & set<string>           & observed            & The set of UI commands.                                                                                               \\
        \hline
        $rm$               & $[0, 1]_{c \times c}$ & observed            & The rank method parameterized by a matrix.                                                                                                                      \\
        \bottomrule
    \end{tabularx}
    \label{tab:simulation-parameters}
\end{table*}

Some parameters are independent of the \QAforML process and are marked ``external''.
For example, before the \QAforML process, there should be a process of applying an ML model to generate a collection of labeled data objects to be checked by the \QAforML process.
Hence, the total number of data objects ($n$) and the confusion matrix ($cm$) of the ML predictions should be known or estimable to UI specialists who conduct the simulation.

UI specialists who are to optimize a \QAforML UI typically have some knowledge about the users who use the UI to perform the \QAforML tasks and intuitions on what design may be appropriate.
We use ``observed'' to categorize parameters to be defined by UI specialists based on their observation of the data objects, users, tasks, and available rank methods and their likely effectiveness in the application context.
UI specialists do not need to fix any parameters to a single value or setting.
The benefit of using simulation is that UI specialists can consider many options.

For example, given some example data objects to be quality-assured, UI specialists can normally judge whether the display size of the data objects in the editing panel is too small or too large.
Meanwhile, UI specialists may be uncertain whether a $4 \times 5$ layout is better than a $5 \times 7$ layout.
UI specialists can simulate both scenarios by assigning different values to the parameter for batch size ($n_{batch})$.
Similarly, UI specialists may consider having users with different \QAforML experiences and at different skill levels.
UI specialists can simulate different options by assigning different values to those user-related parameters ($u_{sl}$ and $u_{su}$).
Choosing parameter values related to the rank methods requires more technical knowledge.
UI specialists may select a few relatively effective rank methods and determine their selection matrices ($rm$).

Table~\ref{tab:simulation-parameters} lists the operator time costs.
Some of them, such as $t_{new}$, are relatively generic.
They are relatively easy to measure and can be reused across different \QAforML applications.
Others, such as $t_{view}$, are highly application-dependent and need to be modeled for each application separately.
We recommend that UI specialists conduct experiments to capture a few trials of using a \QAforML UI.
The timing parameters can be estimated based on captured data one by one, starting from simple ones, such as $t_{new}$.

While we focus on the quality assurance of classification labels, this approach can be adapted to other quality assurance tasks for model predictions (e.g., point cloud segmentation, time series forecasting) as long as the user routine can be modeled.

  \section{Discussion}

Our simulation approach provides an alternative to efficiently evaluate and compare design variations of quality assurance interfaces.
As our approach can be seen as applying the Keystroke-Level Model to quality assurance tasks in grid-based interfaces, we bear the limitations of the Keystroke-Level Model.
The following reflects on the limitations of our approach and future work directions.

\textbf{Relaxing assumptions:}
Our modeling utilizes several assumptions, including that the user makes no mistakes, the machine is fast enough, and the user operations are sequential.
Future work may extend the simulation models to relax these assumptions.
By assuming the user makes no mistakes in carrying out the operations, we estimate a lower bound of the task completion time.
In the future, one may model the time cost caused by users' erroneous use of commands in the interface and the effort to correct such errors.
Latency of the machine operations and parallelized human and machine operations may be considered when estimating the task completion time.
Modeling other factors may also improve the task completion time's estimation accuracy, such as user fatigue that leads to time-varying performance.

\textbf{Using evaluation metrics other than task completion time:}
This work focuses on the evaluation metric of task completion time.
The reason is that \QAforML processes are typically laborious for the user, making the task completion time a critical metric in this scenario.
The goal of quality assurance is well-defined, i.e., fixing incorrect labels, making it feasible to estimate the task completion time.
There are other crucial criteria in evaluating user interfaces, such as user satisfaction and difficulties in learning the UI.
Prior work has studied using model-based evaluation for criteria beyond task completion time, such as using GOMS to evaluate learning time~\cite{John1996GOMS}.
Future work may examine how to adapt prior work to \QAforML.

\textbf{Simulating interface designs beyond the grid-based template:}
Our modeling mainly concerns the grid-based interfaces.
These interfaces are based on buttons and menus, and the number of states of the interface is not too many.
For \QAforML interfaces with interaction mechanisms beyond button-based interactions, such as a paint-based interface to quality-assure image mask predictions, the modeling and simulation of user operations will need more considerations.

\textbf{Validating with more experiment trials from different users:}
In modeling operator time costs in Section~\ref{sec:factor-layout-application-estimation}, we use the experiment log from self-experimentation to demonstrate how to do the modeling.
The implications of simulation results, such as the optimal layout, can be validated through future experiment logs.
By gathering data from multiple users, we may investigate additional user-related aspects, such as the variation of the operator time costs among users.
Note that future experiments are for accumulating records to validate our simulation model's prediction accuracy on the task completion time.
In production scenarios, it is unnecessary and undesirable for evaluators using our model to use user studies to validate all the simulation results.
This would diminish the purpose of using model-based evaluation as an efficient evaluation approach.

\textbf{Improving the simulation approach as a continuous process:}
After optimizing a user interface with the simulation approach and deploying it in practice, it is possible to collect deployment data and compare them with the predictions made by a simulation model.
The comparison may enable the identification of causes of any inconsistency between model predictions and empirical data, therefore facilitating model improvements.
Although many \QAforML user interfaces may be single-use (e.g., for a specific \QAforML workflow), the insights obtained may still enhance the simulation approach for optimizing other \QAforML interfaces in the future.

\textbf{Evaluating the overall cost-benefit of the simulation approach:}
As a medium-to-long-term research agenda, it is highly desirable to collect deployment data from different \QAforML applications to evaluate the overall cost-benefit of using the simulation approach to optimize \QAforML interfaces.
Such a meta-evaluation~\cite{Scriven1969Introduction,Stufflebeam1978Meta}, which has been used in prior HCI research~\cite{Lewis1990Testing,Gong1994Validation}, will allow us to obtain more general findings that apply to a broad range of applications where traditional user studies may be too difficult or costly to conduct.

  \section{Conclusion}

In quality-sensitive applications of \QAforML, the user needs to use \QAforML interfaces intensively.
In such scenarios, interface evaluation and optimization bring significant benefits in terms of saved time for the user.

This work uses model-based evaluation to assess and optimize \QAforML interface design parameters through simulations.
We focus on evaluating the time costs in \QAforML as \QAforML is typically labor-intensive.
Our approach encompasses modeling the user's routine operations in \QAforML \workflow{}s, the user's operation time costs, the algorithmic assistance provided by the interface, and the user's strategy of using the interface with multiple commands available.
We use data-driven modeling to estimate data- and interface-dependent operation time costs.
We demonstrate that through modeling and simulating the task completion time, \QAforML interfaces can be evaluated and optimized.

Using \QAforML application scenarios, including data \extract{}ion from historical visualizations, we demonstrate the need for such an approach and the practical feasibility of using simulation to optimize the interface design.
We demonstrate the influence of various factors on the total time cost of \QAforML, including interface layout, application scenario, availability of interface functions, user's label strategy, default label accuracy, and rank method's precision.
The simulations have derived various findings, such as the dependency of the optimal layout condition on the application.

Model-based evaluation cannot replace user-centered approaches in designing and evaluating interfaces.
It is impractical to expect the modeling approach to capture all the aspects of human factors.
For ill-defined user tasks, such as open-ended data exploration, the evaluator may be unable to model the user's routine operations, making the simulation infeasible.

When the user routine can be modeled, model-based evaluation is a cost-effective alternative for UI specialists to evaluate and optimize the interface.
Simulation makes it possible to swiftly explore potential design parameters of interest and identify the influence of various factors.
It enables fast prototyping and iteration in early-stage interface development before expensive and time-consuming user studies are conducted.
With the simulations that can address many small but potentially costly design issues, we can focus our engagement with users on big design questions about data, ML models, \workflow{}s, and application contexts.

  \section*{Appendices}

Two appendices accompany this work.
The first appendix is titled ``Estimating Operator Time Costs''
\ifx\hideappendix\undefined
    (Appendix~\ref{app:timing-factor}).
\else
    (Appendix A).
\fi
The second appendix is titled ``Modeling and Reestimating Operator Time Costs''
\ifx\hideappendix\undefined
    (Appendix~\ref{app:reestimation}).
\else
    (Appendix B).
\fi
Both appendices are included in the supplementary materials.

\section*{Acknowledgement}

This work has been made possible by the Network of European Data Scientists (NeEDS), a Research and Innovation Staff Exchange (RISE) project under the Marie Skłodowska-Curie Program.
We want to express our gratitude to the people who facilitated this project, particularly Dolores Romero Morales from Copenhagen Business School.

  \bibliographystyle{style/ACM-Reference-Format}
  \bibliography{assets/bibs/papers.bib}
\fi

\ifx\hideappendix\undefined
  \ifx\hidemain\undefined
  \else
    \pagestyle{plain}
  \fi

  \pagebreak
  \appendix
  \noindent\LARGE
\textbf{Appendix to:}

\noindent\LARGE%
\textbf{Simulation-Based Optimization of User Interfaces for\\ Quality-Assuring Machine Learning Model Predictions}\\

\noindent\large
Yu Zhang, University of Oxford\\
Martijn Tennekes, Statistics Netherlands\\
Tim de Jong, Statistics Netherlands\\
Lyana Curier, Open University of the Netherlands\\
Bob Coecke, University of Oxford\\
Min Chen, University of Oxford\\

\normalsize

\section{Estimating Operator Time Costs}
\label{app:timing-factor}

\subsection{Experiment}

We aim to estimate $t_{new}$, $t_{view}$, and $t_{single}$ as outlined in
\ifx\hidemain\undefined
    Section~\ref{sec:factor-layout-application-estimation}.
\else
    Section 5.3 in the main text.
\fi
To facilitate the estimation, we need to gather observations of $<T_{round}, N_{new}, N_{view}, N_{single}>$ through the experiment.
We vary $N_{new}$, $N_{view}$, $N_{single}$ in experiment trials and measure time cost $ T_{round} $.
$N_{new}$ is the number of times the user activate the ``new batch'' command to request a new batch of data objects to quality-assure.
$N_{view}$ is the number of times the user carries out the ``viewing'' action to comprehend the visual representation of a single data object.
$N_{single}$ is the number of times the user activates the ``single edit'' command to change the label of a data object.

\begin{table}[htbp]
    \centering
    \footnotesize
    \caption{
        Experiment design:
        The number of repeated trials conducted for each combination of the independent variables: application, grid layout, number of data objects.
    }
    \begin{tabular}{ccrr}
        \toprule
        \textbf{Application} & \textbf{Layout} & \textbf{\#Objects} & \textbf{\#Trials} \\
        \midrule
        JS-block             & 1 $\times$ 1    & 0                  & 1                 \\
        JS-block             & 1 $\times$ 1    & 1                  & 20                \\
        JS-block             & 1 $\times$ 2    & 0                  & 1                 \\
        JS-block             & 1 $\times$ 2    & 1                  & 15                \\
        JS-block             & 1 $\times$ 2    & 2                  & 15                \\
        JS-block             & 2 $\times$ 2    & 0                  & 1                 \\
        JS-block             & 2 $\times$ 2    & 2                  & 6                 \\
        JS-block             & 2 $\times$ 2    & 4                  & 15                \\
        JS-block             & 2 $\times$ 3    & 0                  & 1                 \\
        JS-block             & 2 $\times$ 3    & 3                  & 10                \\
        JS-block             & 2 $\times$ 3    & 6                  & 10                \\
        JS-block             & 3 $\times$ 4    & 0                  & 1                 \\
        JS-block             & 3 $\times$ 4    & 4                  & 10                \\
        JS-block             & 3 $\times$ 4    & 8                  & 10                \\
        JS-block             & 3 $\times$ 4    & 12                 & 10                \\
        JS-block             & 4 $\times$ 5    & 0                  & 1                 \\
        JS-block             & 4 $\times$ 5    & 10                 & 4                 \\
        JS-block             & 4 $\times$ 5    & 15                 & 6                 \\
        JS-block             & 4 $\times$ 5    & 20                 & 6                 \\
        JS-block             & 6 $\times$ 6    & 0                  & 1                 \\
        JS-block             & 6 $\times$ 6    & 12                 & 4                 \\
        JS-block             & 6 $\times$ 6    & 18                 & 3                 \\
        JS-block             & 6 $\times$ 6    & 24                 & 5                 \\
        JS-block             & 6 $\times$ 6    & 30                 & 6                 \\
        JS-block             & 6 $\times$ 6    & 36                 & 7                 \\
        JS-block             & 8 $\times$ 8    & 0                  & 1                 \\
        JS-block             & 8 $\times$ 8    & 32                 & 4                 \\
        JS-block             & 8 $\times$ 8    & 48                 & 5                 \\
        JS-block             & 8 $\times$ 8    & 64                 & 3                 \\
        \bottomrule
    \end{tabular}%
    \label{tab:experiment-design-a}%
\end{table}%

\begin{table}[htbp]
    \centering
    \footnotesize
    \caption{
        Experiment design continued:
        The number of repeated trials conducted for each combination of the independent variables: application, grid layout, number of data objects.
    }
    \begin{tabular}{ccrr}
        \toprule
        \textbf{Application} & \textbf{Layout} & \textbf{\#Objects} & \textbf{\#Trials} \\
        \midrule
        JS-grouping          & 1 $\times$ 1    & 0                  & 1                 \\
        JS-grouping          & 1 $\times$ 1    & 1                  & 20                \\
        JS-grouping          & 2 $\times$ 3    & 0                  & 1                 \\
        JS-grouping          & 2 $\times$ 3    & 3                  & 7                 \\
        JS-grouping          & 2 $\times$ 3    & 6                  & 6                 \\
        JS-grouping          & 3 $\times$ 4    & 0                  & 1                 \\
        JS-grouping          & 3 $\times$ 4    & 4                  & 3                 \\
        JS-grouping          & 3 $\times$ 4    & 8                  & 3                 \\
        JS-grouping          & 3 $\times$ 4    & 12                 & 3                 \\
        JS-grouping          & 4 $\times$ 5    & 0                  & 1                 \\
        JS-grouping          & 4 $\times$ 5    & 10                 & 5                 \\
        JS-grouping          & 4 $\times$ 5    & 15                 & 5                 \\
        JS-grouping          & 4 $\times$ 5    & 20                 & 5                 \\
        JS-grouping          & 6 $\times$ 6    & 0                  & 1                 \\
        JS-grouping          & 6 $\times$ 6    & 12                 & 5                 \\
        JS-grouping          & 6 $\times$ 6    & 24                 & 5                 \\
        JS-grouping          & 6 $\times$ 6    & 36                 & 5                 \\
        JS-grouping          & 8 $\times$ 8    & 0                  & 1                 \\
        JS-grouping          & 8 $\times$ 8    & 24                 & 1                 \\
        JS-grouping          & 8 $\times$ 8    & 32                 & 4                 \\
        JS-grouping          & 8 $\times$ 8    & 48                 & 5                 \\
        JS-grouping          & 8 $\times$ 8    & 64                 & 3                 \\
        \midrule
        SP-image             & 1 $\times$ 1    & 0                  & 1                 \\
        SP-image             & 1 $\times$ 1    & 1                  & 5                 \\
        SP-image             & 2 $\times$ 3    & 0                  & 1                 \\
        SP-image             & 2 $\times$ 3    & 3                  & 4                 \\
        SP-image             & 2 $\times$ 3    & 6                  & 4                 \\
        SP-image             & 3 $\times$ 4    & 0                  & 1                 \\
        SP-image             & 3 $\times$ 4    & 4                  & 4                 \\
        SP-image             & 3 $\times$ 4    & 8                  & 4                 \\
        SP-image             & 3 $\times$ 4    & 12                 & 4                 \\
        SP-image             & 4 $\times$ 5    & 0                  & 1                 \\
        SP-image             & 4 $\times$ 5    & 10                 & 4                 \\
        SP-image             & 4 $\times$ 5    & 15                 & 4                 \\
        SP-image             & 4 $\times$ 5    & 20                 & 4                 \\
        SP-image             & 5 $\times$ 8    & 0                  & 1                 \\
        SP-image             & 5 $\times$ 8    & 16                 & 4                 \\
        SP-image             & 5 $\times$ 8    & 24                 & 4                 \\
        SP-image             & 5 $\times$ 8    & 32                 & 4                 \\
        SP-image             & 5 $\times$ 8    & 40                 & 4                 \\
        \bottomrule
    \end{tabular}%
    \label{tab:experiment-design-b}%
\end{table}%

\subsubsection{Apparatus}

The experiment is conducted on a laptop with a 15-inch 3:2 screen.
The screen resolution is $ 2560 \times 1600 $.
The screen is viewed by the participant from approximately 80 cm away.
The participant used a physical mouse to conduct the tasks.
The mouse sensitivity is 2000 DPI.

The experiment is carried out for three quality assurance interfaces as shown in
\ifx\hidemain\undefined
    Figure~\ref{fig:grid-based-interfaces}(a - c).
\else
    Fig. 2(a - c) in the main text.
\fi
The \QAforML interfaces are all implemented as web applications that run in a browser.
We implemented a browser plugin to log observations of $ T_{round} $.
One interface is for data \extract{}ion from visualization images that requires quality assurance for JS-block and JS-grouping
\ifx\hidemain\undefined
    (see Figure~\ref{fig:application-workflows}(a)).
\else
    (see Fig. 1(a) in the main text).
\fi
The other interface is for solar panel detection from remote sensing images that requires quality assurance for SP-image
\ifx\hidemain\undefined
    (see Figure~\ref{fig:application-workflows}(b)).
\else
    (see Fig. 1(b) in the main text).
\fi
Each quality assurance interface presents a grid panel with each grid cell corresponding to a data object.

\subsubsection{Procedure}

The experiment is carried out through self-experimentation by the author for proof of concept.
As the developer of the interfaces, the user is familiar with all the interface functionalities.
The user has much experience using the interfaces before the experiment and represents the user group with high expertise in using the interfaces.

We carried out in total 332 trials, as shown in Table~\ref{tab:experiment-design-a} and Table~\ref{tab:experiment-design-b} for the three different applications, with different grid layouts and different numbers of grid cells shown in the interfaces.

Before starting each trial, the browser plugin adds a white overlay to the interface and highlights the location of the grid cells.
After the user clicks the overlay, the timer of the trial starts.
The user quality-assures the labels with the provided functionality of the interface.
Once the user clicks the confirm button, the timer stops, and the time cost of the trial is logged.
Each trial is conducted twice to alleviate the impact of the learning effect on the stability of time costs.
The time is logged when the trial is conducted the second time.
To simulate different levels of default label accuracies, we programmatically randomize the default labels of data objects.
For example, for the binary classification case, given an accuracy $acc$, the program randomly selects $round(acc \cdot n_{batch})$ data objects to flip their true labels.

\subsection{Initial Estimations of Operator Time Costs}
\label{sec:experiment-analysis-initial}

\begin{figure}[!htbp]
    \centering
    \includegraphics[width=\linewidth]{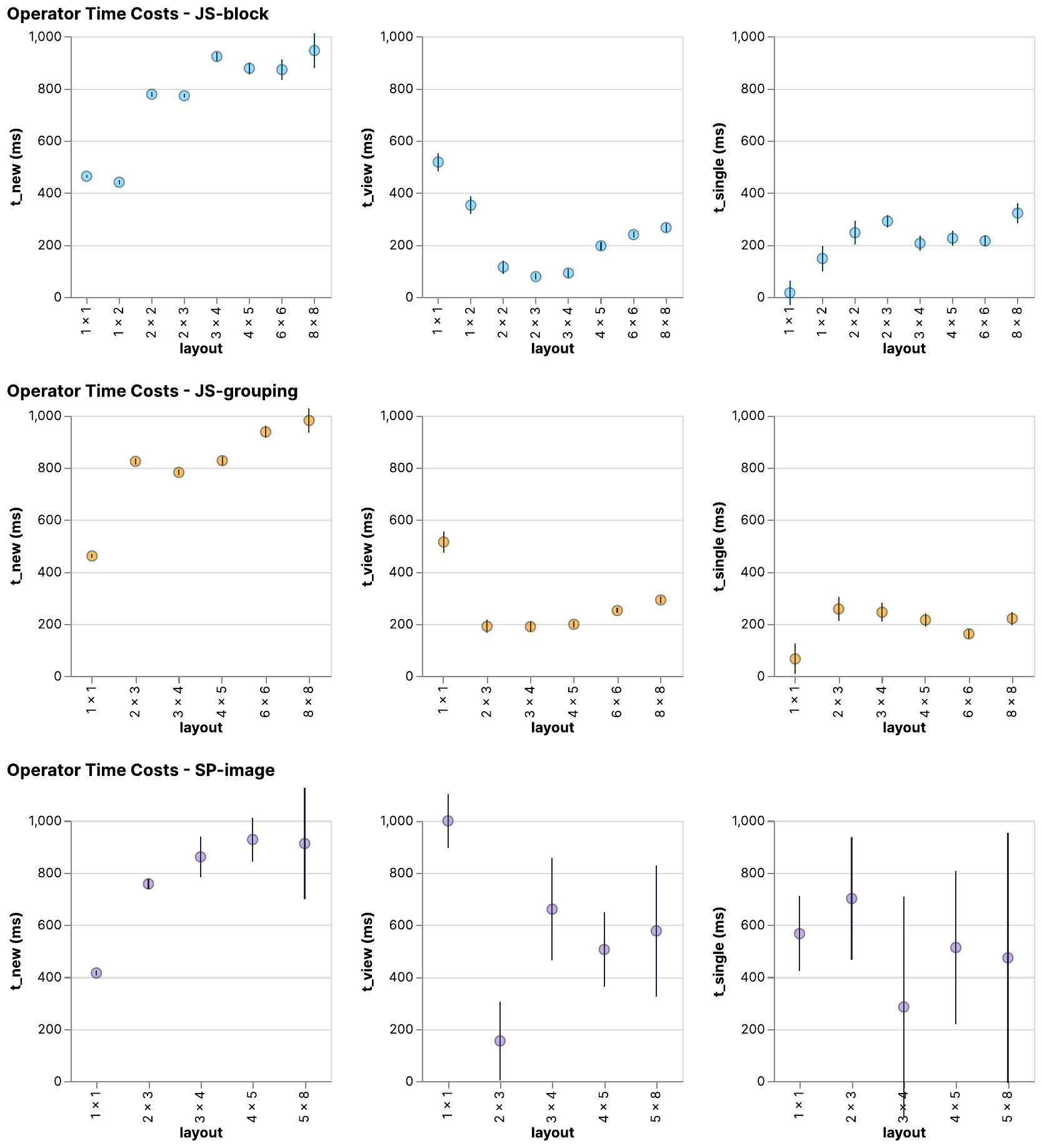}
    \caption{
        Initial estimations of operator time costs:
        The estimated operator time costs $t_{new}$, $t_{view}$, and $t_{single}$ in relation to the grid layouts for all the applications.
        The error bars show the estimated standard error of the estimation.
        As the number of grid cells grows, $t_{new}$ increases, $t_{view}$ first increases then decreases, $t_{single}$ first increases then stays stable.
    }
    \Description[
        Fully described in the caption.
    ]{}
    \label{fig:operator-time-costs-initial}
\end{figure}

Using the experiment data, we estimate $ t_{new} $, $t_{view}$, $t_{single}$ with multiple linear regression.
Specifically, we go through the following estimation procedure for each combination of <application, layout>.
We fit an individual model for each combination of <application, layout> because the application and layout may significantly influence the operation time costs.

Given <application, layout>, assume $ T_{round} $ follows a multiple linear model with regards to $ N_{new} $, $N_{view}$, $N_{single}$ as $ T_{round} = N_{new}t_{new} + N_{view}t_{view} + N_{single}t_{single} $ in
\ifx\hidemain\undefined
    Equation~\ref{eq:basic-sum}.
\else
    Equation 1 in the main text.
\fi
We fit a multiple linear regression model for $ T_{round} $ to solve for $ t_{new} $, $ t_{view} $ and $ t_{single} $.
The standard error, $ R^2 $, and significance of the fitting are computed.
Table~\ref{tab:experiment-summary} and Figure~\ref{fig:operator-time-costs-initial} show the estimation result.

\begin{table}[htbp]
    \centering
    \scriptsize
    \caption{
        Initial estimation of operator time costs:
        The estimated time cost of unit user operations using multiple linear regression.
        For each estimation, the estimated standard error is shown in the parenthesis.
        The statistical significance is marked.
        ** means $ p \leq 0.01 $ and * means $ p \leq 0.05 $.
        $R^2$ columns show the R-squared of the entire linear model $ T_{round} = N_{new} t_{new} + N_{view} t_{view} + N_{single} t_{single} $.
        It can be seen that for all the experimented applications and grid layouts, $ N_{new} $ contributes linearly to $ T_{round} $.
        In most cases, $ N_{view} $ and $ N_{single} $ contributes linearly to $ T_{round} $.
    }
    \begin{tabular}{ccllll}
        \toprule
        \textbf{Application} & \textbf{Layout} & $\mathbf{t_{new}}$ (ms) & $\mathbf{t_{view}}$ (ms) & $\mathbf{t_{single}}$ (ms) & $\mathbf{R^2}$ \\
        \midrule
        JS-block             & 1 $\times$ 1    & 463 ($\pm$ 5) **        & 518 ($\pm$ 35) **        & 16 ($\pm$ 47)              & 0.998          \\
        \midrule
        JS-block             & 1 $\times$ 2    & 441 ($\pm$ 8) **        & 352 ($\pm$ 33) **        & 148 ($\pm$ 49) **          & 0.993          \\
        \midrule
        JS-block             & 2 $\times$ 2    & 778 ($\pm$ 10) **       & 115 ($\pm$ 25) **        & 247 ($\pm$ 46) **          & 0.997          \\
        \midrule
        JS-block             & 2 $\times$ 3    & 773 ($\pm$ 7) **        & 78 ($\pm$ 12) **         & 291 ($\pm$ 24) **          & 0.999          \\
        \midrule
        JS-block             & 3 $\times$ 4    & 923 ($\pm$ 18) **       & 92 ($\pm$ 16) **         & 206 ($\pm$ 28) **          & 0.993          \\
        \midrule
        JS-block             & 4 $\times$ 5    & 878 ($\pm$ 23) **       & 196 ($\pm$ 15) **        & 226 ($\pm$ 28) **          & 0.997          \\
        \midrule
        JS-block             & 6 $\times$ 6    & 873 ($\pm$ 39) **       & 239 ($\pm$ 12) **        & 215 ($\pm$ 21) **          & 0.995          \\
        \midrule
        JS-block             & 8 $\times$ 8    & 946 ($\pm$ 68) **       & 266 ($\pm$ 18) **        & 322 ($\pm$ 38) **          & 0.997          \\
        \midrule
        JS-grouping          & 1 $\times$ 1    & 461 ($\pm$ 7) **        & 515 ($\pm$ 40) **        & 66 ($\pm$ 59)              & 0.997          \\
        \midrule
        JS-grouping          & 2 $\times$ 3    & 824 ($\pm$ 12) **       & 191 ($\pm$ 25) **        & 258 ($\pm$ 45) **          & 0.998          \\
        \midrule
        JS-grouping          & 3 $\times$ 4    & 782 ($\pm$ 10) **       & 190 ($\pm$ 19) **        & 245 ($\pm$ 35) **          & 0.999          \\
        \midrule
        JS-grouping          & 4 $\times$ 5    & 827 ($\pm$ 17) **       & 199 ($\pm$ 14) **        & 215 ($\pm$ 25) **          & 0.998          \\
        \midrule
        JS-grouping          & 6 $\times$ 6    & 937 ($\pm$ 22) **       & 252 ($\pm$ 10) **        & 162 ($\pm$ 18) **          & 0.998          \\
        \midrule
        JS-grouping          & 8 $\times$ 8    & 981 ($\pm$ 46) **       & 292 ($\pm$ 14) **        & 221 ($\pm$ 25) **          & 0.998          \\
        \midrule
        SP-image             & 1 $\times$ 1    & 416 ($\pm$ 9) **        & 999 ($\pm$ 103) **       & 566 ($\pm$ 145) *          & 0.999          \\
        \midrule
        SP-image             & 2 $\times$ 3    & 757 ($\pm$ 20) **       & 155 ($\pm$ 151)          & 701 ($\pm$ 234) *          & 0.997          \\
        \midrule
        SP-image             & 3 $\times$ 4    & 861 ($\pm$ 78) **       & 660 ($\pm$ 198) **       & 285 ($\pm$ 423)            & 0.976          \\
        \midrule
        SP-image             & 4 $\times$ 5    & 928 ($\pm$ 84) **       & 506 ($\pm$ 142) **       & 513 ($\pm$ 295)            & 0.988          \\
        \midrule
        SP-image             & 5 $\times$ 8    & 912 ($\pm$ 214) **      & 577 ($\pm$ 252) *        & 473 ($\pm$ 479)            & 0.976          \\
        \bottomrule
    \end{tabular}
    \label{tab:experiment-summary}
\end{table}

  \section{Modeling and Reestimating Operator Time Costs}
\label{app:reestimation}

In the following, we introduce a process of smoothing the operator time costs estimated in Section~\ref{sec:experiment-analysis-initial}.
We fit curves for the operator time costs and use the values on the curves to substitute the initial estimations of the operator time costs in Section~\ref{sec:experiment-analysis-initial}.

\subsection{Model \texorpdfstring{$\mathbf{t_{new}}$}{t\_new}}

\paragraph{Goal}

Reestimate $t_{new}$ as a function of interface parameter $n_{batch}$ to smooth the estimation.

\paragraph{Procedure}

There are 19 combinations of <application, layout> and thus 19 $<n_{batch}, t_{new}>$ samples.
We fit a model for all three applications combined.
We have used the Nonlinear Regression Tool~\cite{Xuru2006Online} to produce a set of candidate model functions $ t_{new} = f(n_{batch}) $.

\begin{itemize}
    \item With the number of parameters = 2, the top 3 functions are
          \begin{itemize}
              \item $y = \frac{ax}{x + b} \mbox{ with } R^2 = 0.9205$
              \item $y = ae^{b/x} \mbox{ with } R^2 = 0.8989$
              \item $y = a + b/x \mbox{ with } R^2 = 0.8630$
          \end{itemize}

    \item With the number of parameters = 3, the top 3 functions are
          \begin{itemize}
              \item $y = a + b/x + c/x^2 \mbox{ with } R^2 = 0.9381$
              \item $y = ax + \frac{bx}{x + c} \mbox{ with } R^2 = 0.9231$
              \item $y = \frac{1}{ax^b + c} \mbox{ with } R^2 = 0.9225$
          \end{itemize}

    \item With the number of parameters = 4, the top 3 functions are
          \begin{itemize}
              \item $y = a + b/x + c/x^2 + d/x^3 \mbox{ with } R^2 = 0.9486$
              \item $y = a + b/x + c/x^2 + dx \mbox{ with } R^2 = 0.9486$
              \item $y = a + b/x + c/x^2 \mbox{ with } R^2 = 0.9381$
          \end{itemize}
\end{itemize}

It can be seen that $ y = a + b/x + c/x^2 $ is reoccurring.
It is the best when \#parameters = 3 and the third best when \#parameters = 4.
Adding additional terms generate the best function when \#parameters = 4.
Removing the $ c/x^2 $ term generates the third-best when \#parameters = 2.

We refer to the literature for more function options.
The ``new batch'' button clicking relates to the object-pointing task in Fitts' law~\cite{Fitts1992Information}.
Fitts' law suggests that the object-pointing time cost can be modeled as $ T = a + blog(\frac{A}{W} + 1) $.
$ W $ is the size of the target object, $ A $ is the distance from the initial cursor position to the target object, whereas $a$ and $b$ are parameters to be fitted, which may depend on interface and apparatus settings.

To activate the ``new batch'' command, the user needs to move the cursor to the ``new batch'' button and click it.
This procedure is similar to the Fitts' law setting, while a significant difference from typical Fitts' law settings is that the distance to the target (the ``new batch'' button) is not well-defined.
Because the last position of the cursor before the user decides to activate the ``new batch'' command can be an arbitrary point in the interface.

We adopt the following approximation to examine whether Fitts' law works for our scenario.
Assume the aspect ratio of the grid cells does not change with the increase of $n_{batch}$.
Denote the layout as $ xf \times yf $ where $ x, y, f $ are integers.
The maximal distance of points in the interface is thus $ \sqrt{x^2 + y^2}f $.
As $ f $ increases, the number of grid cells $xyf^2$ increases at the rate of $f^2$, and the maximal distance $ \sqrt{x^2 + y^2}f $ increases at the rate of $f$.
We approximate the distance $ A $ in Fitts' law by the maximal distance $ \sqrt{x^2 + y^2}f $ and then represent it as $c \sqrt{n_{batch}}$ where $ c $ is an interface related constant.
Thus, we get $ t_{new} = a + blog_2(c\sqrt{n_{batch}} + 1) $.
However, this function is not convex and is hard to optimize.
Therefore, we modify it to $ t_{new} = a + blog_2(\sqrt{n_{batch}} + 1) $ by removing $c$.

In short, we additionally consider the model function $ t_{new} = a + blog_2(\sqrt{n_{batch}} + 1) $ to see whether Fitts' law applies to our scenario.

Thus, we choose $ y = a + b/x + c/x^2 $ and $ y = a + blog_2(\sqrt{x} + 1) $ to be the candidate function families for $ t_{new} $.
We fit the models with linear regression.

\paragraph{Outcome}

\begin{figure}[htbp]
    \centering
    \includegraphics[width=0.5\linewidth]{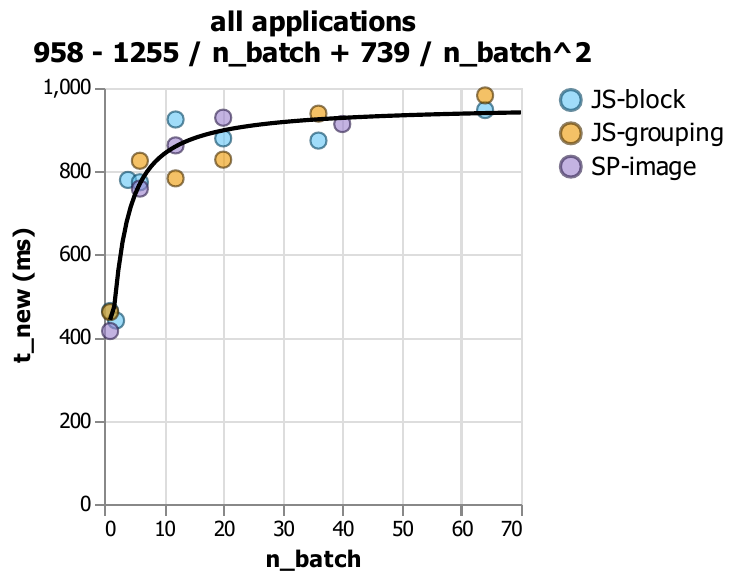}
    \caption{
        The estimated model of $t_{new}$ as a function of $n_{batch}$ for the three applications using $ y = a + b/x + c/x^2 $.
    }
    \Description[
        Fully described in the main text of the section.
    ]{}
    \label{fig:model-t-new}
\end{figure}

\begin{itemize}
    \item For $y = a + b/x + c/x^2$, the function fitted is:
          \begin{align*}
              \begin{aligned}
                   & t_{new} = 958.0085 - \frac{1254.8737}{n_{batch}} + \frac{739.4750}{n_{batch}^2} \\
                   & \mbox{ with adjusted } R^2 = 0.9303, SE = 49.4461
              \end{aligned}
          \end{align*}

    \item For $y = a + blog(\sqrt{x} + 1)$, the function fitted is:
          \begin{align*}
              \begin{aligned}
                   & t_{new} = 280.0475 + 341.5430 \cdot log_2(\sqrt{n_{batch}} + 1) \\
                   & \mbox{ with adjusted } R^2 = 0.8004, SE = 83.6951
              \end{aligned}
          \end{align*}
\end{itemize}

The fitting of $y = a + blog(\sqrt{x} + 1)$ is worse than $y = a + b/x + c/x^2$ in terms of adjusted $R^2$ and $SE$.
It implies that the model suggested by Fitts' law is not numerically accurate in this scenario.
Thus, we decide to adopt the model in Figure~\ref{fig:model-t-new}:
\begin{equation}
    \mathbf{t_{new} = 958.0085 - \frac{1254.8737}{n_{batch}} + \frac{739.4750}{n_{batch}^2}}
\end{equation}

\subsection{Reestimate \texorpdfstring{$\mathbf{t_{view}}$}{t\_view} and \texorpdfstring{$\mathbf{t_{single}}$}{t\_single} by \texorpdfstring{$\mathbf{t_{new}}$}{t\_new} Model}

\paragraph{Goal}

Reestimate $ t_{view} $ and $ t_{single} $ by removing the previous modeled $t_{new}$ from the equations to possibly reduce the noise and make sure the original equations still approximately hold.

\paragraph{Procedure}

With the $t_{new}$ modeled in the last section, we reestimate $t_{view}$ and $t_{single}$ from the experiment data by removing the contribution of $t_{new}$ as

\begin{gather}
    \begin{bmatrix}
        t_{view} \\ t_{single}
    \end{bmatrix}
    = (X^T X)^{-1} X^T
    \begin{bmatrix}
        T_1 - N_{new, 1}t_{new} \\ T_2 - N_{new, 2}t_{new} \\ ... \\ T_{n} - N_{new, n}t_{new}
    \end{bmatrix}
\end{gather}
where X is the $N_{view}$ and $N_{single}$ measured for the $ n $ trials
\begin{gather*}
    X =
    \begin{bmatrix}
        N_{view, 1} & N_{single, 1} \\
        N_{view, 2} & N_{single, 2} \\
        ...         & ...           \\
        N_{view, n} & N_{single, n} \\
    \end{bmatrix}
\end{gather*}
and the $t_{new}$ in the formula use modeled $t_{new}$ values in the last section.

\paragraph{Outcome}

\begin{figure}[htbp]
    \centering
    \includegraphics[width=\linewidth]{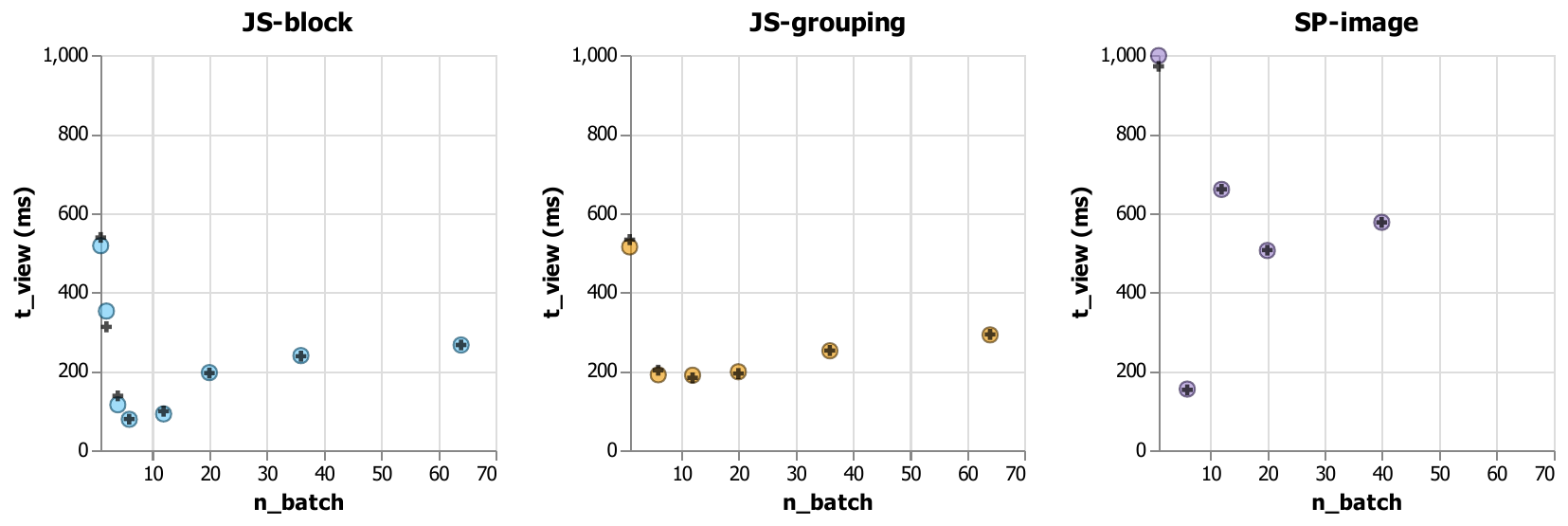}
    \caption{
        $t_{view}$ reestimation by $t_{new}$:
        The influence of reestimation to $t_{view}$ as a function of $n_{batch}$ for the three applications.
        The colored dots are initial estimations by multiple linear regression.
        The black crosses are reestimations by putting $ t_{new} $ back.
    }
    \Description[
        Fully described in the caption.
    ]{}
    \label{fig:t-view-reestimate-1}
\end{figure}

\begin{figure}[htbp]
    \centering
    \includegraphics[width=\linewidth]{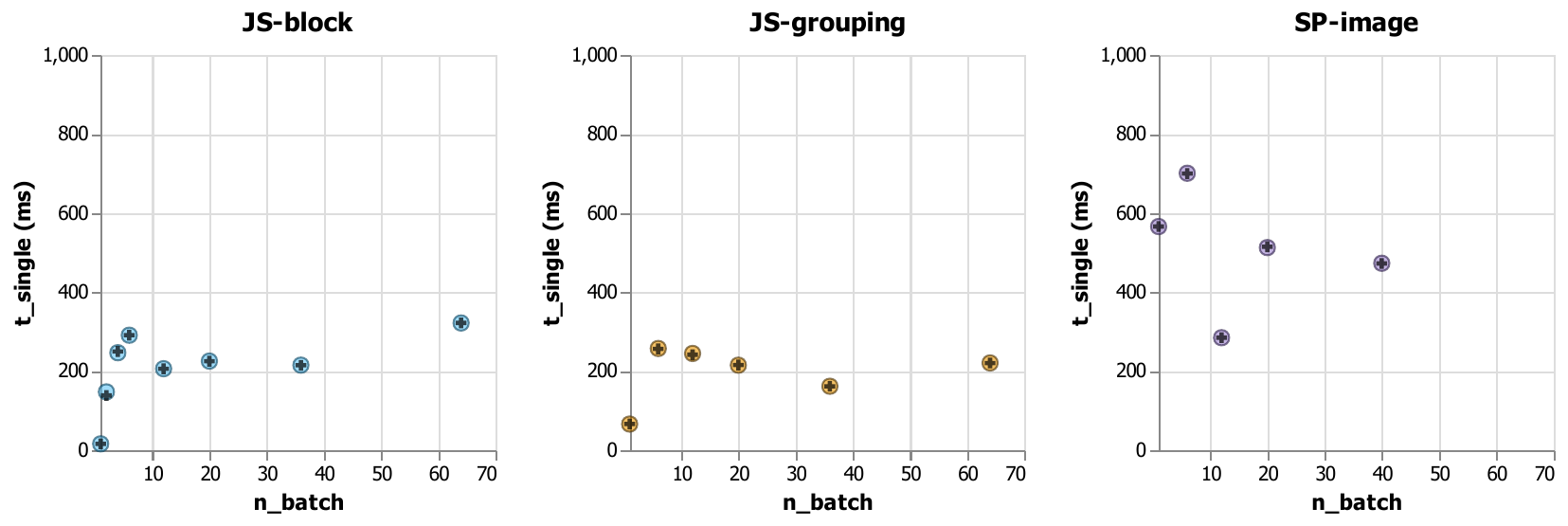}
    \caption{
        $t_{single}$ reestimation by $t_{new}$:
        The influence of reestimation to $t_{single}$ as a function of $n_{batch}$ for the three applications.
        The colored dots are initial estimations by multiple linear regression.
        The black crosses are reestimations by putting $ t_{new} $ back.
    }
    \Description[
        Fully described in the caption.
    ]{}
    \label{fig:t-single-preprocess}
\end{figure}

Figure~\ref{fig:t-view-reestimate-1} and Figure~\ref{fig:t-single-preprocess} show the influence of the reestimation on $t_{view}$ and $t_{single}$.

\paragraph{Comments}

For both $ t_{view} $ and $ t_{single} $, reestimation makes little change to the data points.

\begin{itemize}
    \item \textbf{For} $\mathbf{t_{view}}$:
          For JS-block and JS-grouping, $ t_{view} $ seems to follow a U-shaped curve.
          For SP-image, $ t_{view} $  is noisy and hard to interpret.
          The pattern of $ t_{view} $ is complex.
          We decide the leave $ t_{view} $ at the moment.

    \item \textbf{For} $\mathbf{t_{single}}$:
          For JS-block and JS-grouping, $t_{single}$ exhibits a constant or weak linear pattern except for the first data point.
\end{itemize}

The actions of single edit may happen during and after viewing actions.
The attribution of time to single edit and view may be less accurate when few data objects are sampled (i.e., when $n_{batch}$ is small), and the number of single edit actions is sparse.

Compared to the data points' pattern in JS-block and JS-grouping, the second and third data points of $t_{single}$ of SP-image appear anomalous.
For the SP-image example, if we ignore the second and third outlying data points, the remaining data points exhibit a constant or weak linear pattern.

It is reasonable to assume that the time cost of a single edit action ($t_{single}$) is reasonably consistent within an application but inconsistent across different applications.
Figure~\ref{fig:t-single-preprocess} evidences this while showing that it has a less complicated pattern than $t_{view}$.

The pattern of $t_{single}$ is more straightforward than $t_{view}$.
Thus, we decide to model it first.

\subsection{Model Reestimated \texorpdfstring{$\mathbf{t_{single}}$}{t\_single}}

\paragraph{Goal}

Model the reestimated $ t_{single} $ to reduce noise.

\paragraph{Procedure}

We model $t_{single}$ as a function of $n_{batch}$.
There is no clear evidence that $t_{single}$ is application-independent.
Thus, we fit a model for each application.
For JS-block, there are 8 data points.
For JS-grouping, there are 6 data points.
For SP-image, there are 5 data points.
We have used the Nonlinear Regression Tool~\cite{Xuru2006Online} to produce a set of candidate model functions $ t_{single} = f(n_{batch}) $.

\begin{itemize}
    \item With the number of parameters = 2, the top 3 functions are
          \begin{itemize}
              \item for JS-block:
                    \begin{itemize}
                        \item $y = a + b/x \mbox{ with } R^2 = 0.7795$
                        \item $y = ae^{b/x} \mbox{ with } R^2 = 0.7049$
                        \item $y = \frac{ax}{x + b} \mbox{ with } R^2 = 0.6303$
                    \end{itemize}

              \item for JS-grouping:
                    \begin{itemize}
                        \item $y = a + b/x \mbox{ with } R^2 = 0.6940$
                        \item $y = ae^{b/x} \mbox{ with } R^2 = 0.6400$
                        \item $y = ax^{b/x} \mbox{ with } R^2 = 0.5938$
                    \end{itemize}

              \item for SP-image:
                    \begin{itemize}
                        \item $y = acos(bx) \mbox{ with } R^2 = 0.3560$
                        \item $y = a + ln(x) \mbox{ with } R^2 = 0.1385$
                        \item $y = ax^b \mbox{ with } R^2 = 0.1364$
                    \end{itemize}
          \end{itemize}

    \item With the number of parameters = 3, the top 3 functions are
          \begin{itemize}
              \item for JS-block:
                    \begin{itemize}
                        \item $y = \frac{1}{a + bx^c} \mbox{ with } R^2 = 0.8310$
                        \item $y = a + bx + c/x^2 \mbox{ with } R^2 = 0.8240$
                        \item $y = a + b/x + c/x^2 \mbox{ with } R^2 = 0.7939$
                    \end{itemize}

              \item for JS-grouping:
                    \begin{itemize}
                        \item $y = a + b/x + c/x^2 \mbox{ with } R^2 = 0.9018$
                        \item $y = ae^{b/{x + c}} \mbox{ with } R^2 = 0.8927$
                        \item $y = ae^{b/x + cln(x)} \mbox{ with } R^2 = 0.8506$
                    \end{itemize}

              \item for SP-image:
                    \begin{itemize}
                        \item $y = ax^{bx^c} \mbox{ with } R^2 = 0.4980$
                        \item $y = acos(bx) \mbox{ with } R^2 = 0.3560$
                        \item $y = a + b/x + c/x^2 \mbox{ with } R^2 = 0.3172$
                    \end{itemize}
          \end{itemize}

    \item With the number of parameters = 4, the top 3 functions are
          \begin{itemize}
              \item for JS-block:
                    \begin{itemize}
                        \item $y = a + bx + c/x + dln(x) \mbox{ with } R^2 = 0.9059$
                        \item $y = a + bx^{0.5} + cx + dx^{1.5} \mbox{ with } R^2 = 0.8574$
                        \item $y = a + bx + ((bx - c)^2 - d)^{0.5} \mbox{ with } R^2 = 0.8332$
                    \end{itemize}

              \item for JS-grouping:
                    \begin{itemize}
                        \item $y = acos(x + b) - ccos(2x + b) - dcos(3x + b) \mbox{ with } R^2 = 0.9957$
                        \item $y = a + b^{0.5} + cx + dx^{1.5} \mbox{ with } R^2 = 0.9804$
                        \item $y = a + bx + c/x + dln(x) \mbox{ with } R^2 = 0.9562$
                    \end{itemize}

              \item for SP-image:
                    \begin{itemize}
                        \item $y = a + b/x + c/x^2 + d/x^3 \mbox{ with } R^2 = 0.7832$
                        \item $y = a + bsin(cx + d) \mbox{ with } R^2 = 0.6659$
                        \item $y = a + bx + c/x + d/x^2 \mbox{ with } R^2 = 0.5589$
                    \end{itemize}
          \end{itemize}
\end{itemize}

It can be seen that $ y = a + b/x + c/x^2 $ is reoccurring.
It is the third-best for JS-block and SP-image and best for JS-grouping when \#parameters = 3.
Removing the $ c/x^2 $ term generates the best for JS-block and JS-grouping when \#parameters = 2.
Adding the $ d/x^3 $ term generates the best for SP-image when \#parameters = 4.

Thus, we choose $ y = a + b/x + c/x^2 $ as a candidate function family for $ t_{single} $.

Moreover, we noticed that if we ignore the $ 1 \times 1 $ layout for JS-block and JS-grouping, $ t_{single} $ seems to follow a constant or weak linear pattern.
Thus, we also choose $ y = a $ and $ y = a + bx $ to be candidate function families for $ t_{single} $

We refer to the literature for more function options.
The single edit button clicking is somehow related to the decision-making task in Hick's law~\cite{Hick1952Rate}.
Hick's law suggests that the average reaction time to choose among $n$ equally probable choices can be modeled as $ T = blog_2(n + 1) $ where $ b $ is a parameter to be fitted.
If we assume the solved $t_{single}$ is a time cost for the user to point to a grid cell and click it, $t_{single}$ may be modeled as $t_{single} = a + blog_2(n_{batch} + 1)$.
The $ a $ term denotes the constant cost of clicking, while the $ blog_2(n_{batch} + 1) $ term denotes the decision time following Hick's law.
Our assumption needs further investigation, as the decision time may be included in $t_{view}$ instead of $t_{single}$.

In summary, we fit 4 models with linear regression: $ y = a + b/x + c/x^2 $, $ y = a + blog_2(x + 1) $, $ y = a $, $ y = a + bx $.

\paragraph{Outcome}

\begin{itemize}
    \begin{figure}[htbp]
        \centering
        \includegraphics[width=\linewidth]{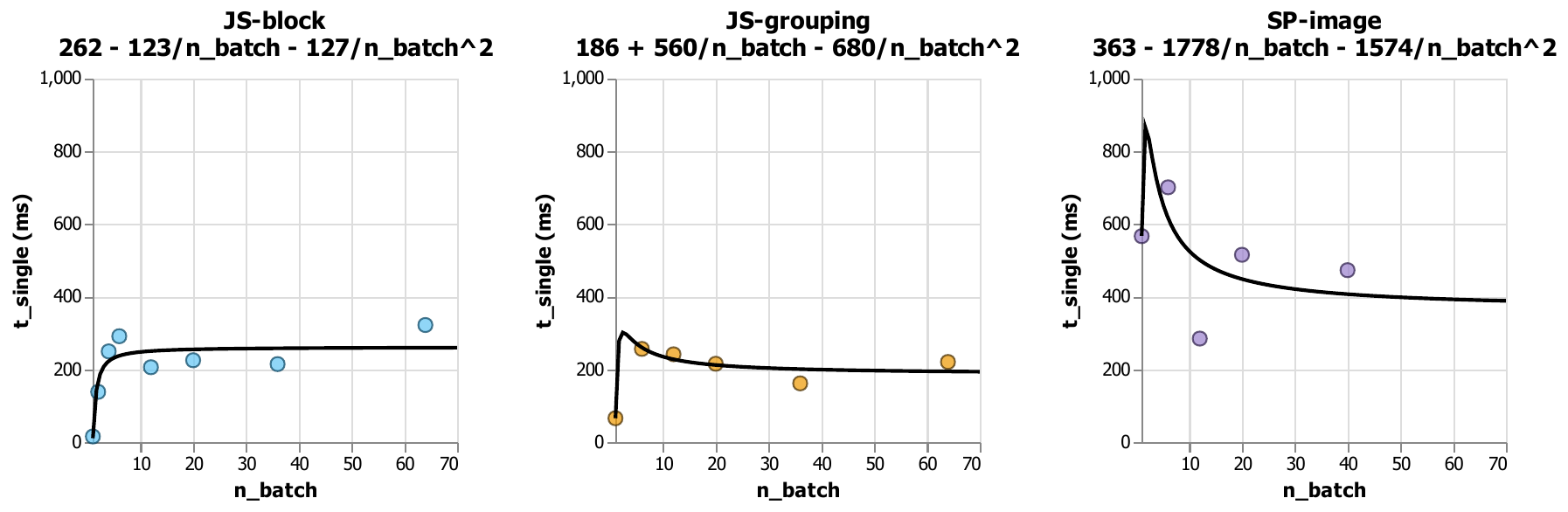}
        \caption{
            The fitted $t_{single}$ as a function of $n_{batch}$ for JS-block, JS-grouping, and SP-image using $y = a + b/x + c/x^2$.
        }
        \Description[
            Fully described in the main text of the section.
        ]{}
        \label{fig:model-t-single-a-b/x-c/x^2}
    \end{figure}

    \item \textbf{For} $\mathbf{y = a + b/x + c/x^2}$:
          Figure~\ref{fig:model-t-single-a-b/x-c/x^2} shows the fitting result for $ t_{single} $.
          For SP-image, due to the noisy data, the fitted model contains a spike.
          The functions fitted are:
          \begin{itemize}
              \item for JS-block:
                    \begin{align*}
                        \begin{aligned}
                             & t_{single} = 261.8083 - 123.2102 / n_{batch} - 127.2516 / n_{batch}^2 \\
                             & \mbox{ with adjusted } R^2 = 0.7115, SE = 51.2691
                        \end{aligned}
                    \end{align*}

              \item for JS-grouping:
                    \begin{align*}
                        \begin{aligned}
                             & t_{single} = 186.0985 + 560.1921/n_{batch} - 680.1595/n_{batch}^2 \\
                             & \mbox{ with adjusted } R^2 = 0.8363, SE = 28.4728
                        \end{aligned}
                    \end{align*}

              \item for SP-image:
                    \begin{align*}
                        \begin{aligned}
                             & t_{single} = 363.5150 - 1778.1390/n_{batch} - 1574.2689/n_{batch}^2 \\
                             & \mbox{ with adjusted } R^2 = -0.3656, SE = 176.7216
                        \end{aligned}
                    \end{align*}
          \end{itemize}

    \item \textbf{For} $\mathbf{y = a + blog_2(x + 1)}$:
          The functions fitted are:
          \begin{itemize}
              \item for JS-block:
                    \begin{align*}
                        \begin{aligned}
                             & t_{single} = 80.7479 + 54.3526 log_2(n_{batch} + 1) \\
                             & \mbox{ with adjusted } R^2 = 0.3991, SE = 73.9862
                        \end{aligned}
                    \end{align*}

              \item for JS-grouping:
                    \begin{align*}
                        \begin{aligned}
                             & t_{single} = 117.6121 + 28.5219 log_2(n_{batch} + 1) \\
                             & \mbox{ with adjusted } R^2 = 0.0675, SE = 67.9599
                        \end{aligned}
                    \end{align*}

              \item for SP-image:
                    \begin{align*}
                        \begin{aligned}
                             & t_{single} = 628.0140 - 50.1844 log_2(n_{batch} + 1) \\
                             & \mbox{ with adjusted } R^2 = -0.1393, SE = 161.4152
                        \end{aligned}
                    \end{align*}
          \end{itemize}

          \begin{figure}[htbp]
              \centering
              \includegraphics[width=\linewidth]{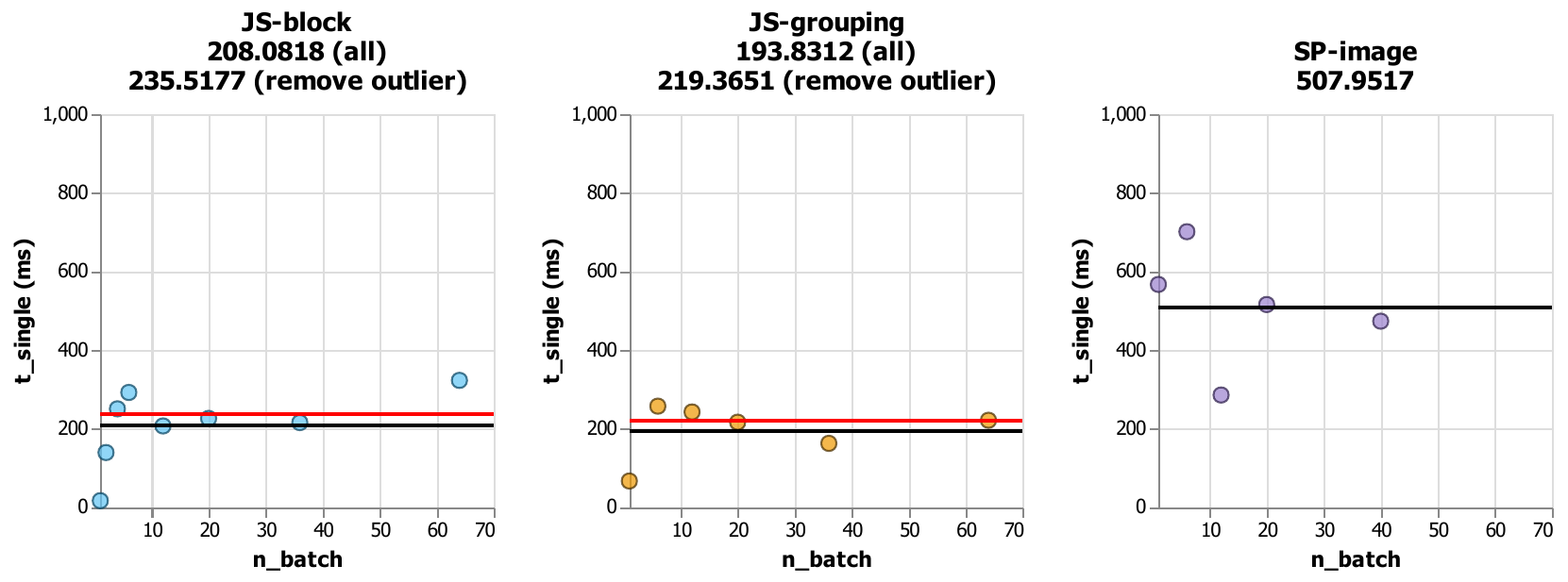}
              \caption{
                  The fitted $t_{single}$ as a function of $n_{batch}$ for JS-block, JS-grouping, and SP-image using $y = a$.
                  The black line corresponds to the estimation without removing the outlier.
                  The red line corresponds to the estimation after removing the outlier.
              }
              \Description[
                  Fully described in the main text of the section.
              ]{}
              \label{fig:model-t-single-a}
          \end{figure}

    \item \textbf{For} $\mathbf{y = a}$:
          Figure~\ref{fig:model-t-single-a} shows the fitting result for $ t_{single} $.
          The functions fitted are:
          \begin{itemize}
              \item for JS-block:
                    \begin{itemize}
                        \item If we use all the data points:
                              \begin{align*}
                                  \begin{aligned}
                                       & t_{single} = 208.0818                        \\
                                       & \mbox{ with adjusted } R^2 = 0, SE = 95.4480
                                  \end{aligned}
                              \end{align*}

                        \item If we remove the first data point, which seems an outlier:
                              \begin{align*}
                                  \begin{aligned}
                                       & t_{single} = 235.5177                        \\
                                       & \mbox{ with adjusted } R^2 = 0, SE = 60.0269
                                  \end{aligned}
                              \end{align*}
                    \end{itemize}

              \item for JS-grouping:
                    \begin{itemize}
                        \item If we use all the data points:
                              \begin{align*}
                                  \begin{aligned}
                                       & t_{single} = 193.8312                        \\
                                       & \mbox{ with adjusted } R^2 = 0, SE = 70.3764
                                  \end{aligned}
                              \end{align*}

                        \item If we remove the first data point, which seems an outlier:
                              \begin{align*}
                                  \begin{aligned}
                                       & t_{single} = 219.3651                        \\
                                       & \mbox{ with adjusted } R^2 = 0, SE = 36.0720
                                  \end{aligned}
                              \end{align*}
                    \end{itemize}

              \item for SP-image:
                    \begin{align*}
                        \begin{aligned}
                             & t_{single} = 507.9517                         \\
                             & \mbox{ with adjusted } R^2 = 0, SE = 151.2289
                        \end{aligned}
                    \end{align*}
          \end{itemize}

          \begin{figure}[htbp]
              \centering
              \includegraphics[width=\linewidth]{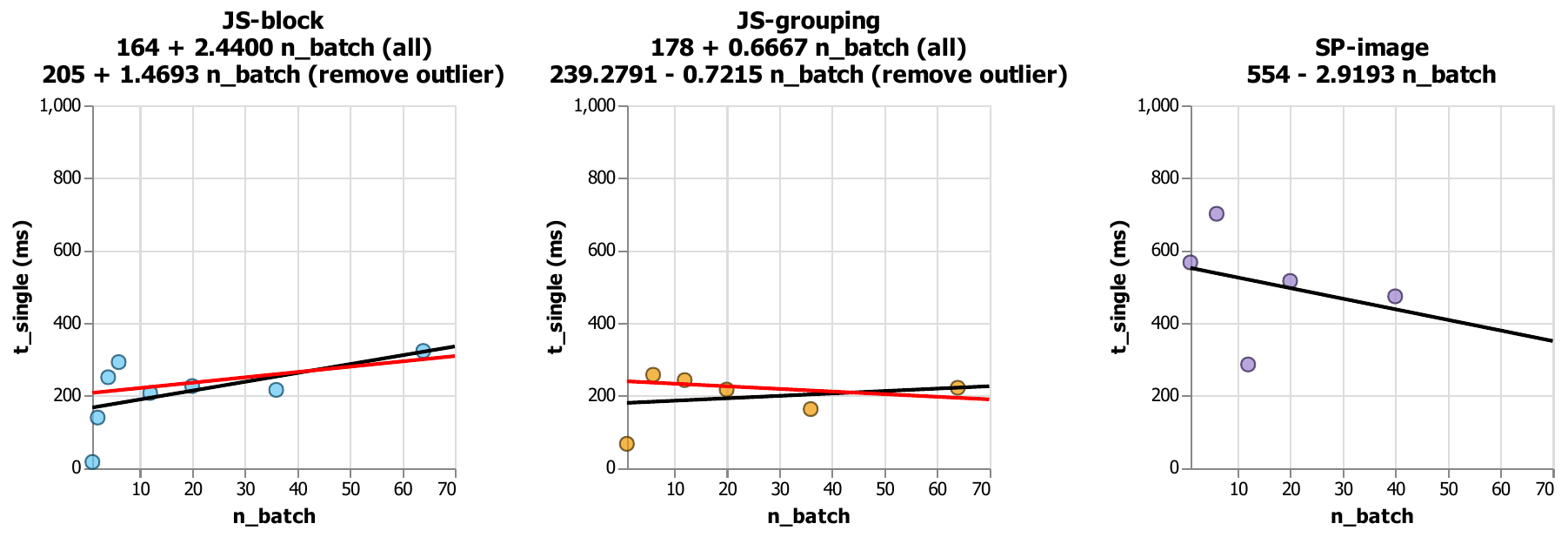}
              \caption{
                  The fitted $t_{single}$ as a function of $n_{batch}$ for JS-block, JS-grouping, and SP-image using $y = a + bx$.
                  The black line corresponds to the estimation without removing the outlier.
                  The red line corresponds to the estimation after removing the outlier.
              }
              \Description[
                  Fully described in the main text of the section.
              ]{}
              \label{fig:model-t-single-a-bx}
          \end{figure}

    \item \textbf{For} $\mathbf{y = a + bx}$:
          Figure~\ref{fig:model-t-single-a-bx} shows the fitting result for $ t_{single} $.
          The functions fitted are:
          \begin{itemize}
              \item for JS-block:
                    \begin{itemize}
                        \item If we use all the data points:
                              \begin{align*}
                                  \begin{aligned}
                                       & t_{single} = 163.8567 + 2.4400 n_{batch}          \\
                                       & \mbox{ with adjusted } R^2 = 0.1998, SE = 85.3806
                                  \end{aligned}
                              \end{align*}

                        \item If we remove the first data point, which seems an outlier:
                              \begin{align*}
                                  \begin{aligned}
                                       & t_{single} = 205.2919 + 1.4693 n_{batch}          \\
                                       & \mbox{ with adjusted } R^2 = 0.1630, SE = 54.9157
                                  \end{aligned}
                              \end{align*}
                    \end{itemize}

              \item for JS-grouping:
                    \begin{itemize}
                        \item If we use all the data points:
                              \begin{align*}
                                  \begin{aligned}
                                       & t_{single} = 178.3868 + 0.6667 n_{batch}           \\
                                       & \mbox{ with adjusted } R^2 = -0.1882, SE = 76.7149
                                  \end{aligned}
                              \end{align*}

                        \item If we remove the first data point, which seems an outlier:
                              \begin{align*}
                                  \begin{aligned}
                                       & t_{single} = 239.2791 - 0.7215 n_{batch}           \\
                                       & \mbox{ with adjusted } R^2 = -0.0448, SE = 36.8718
                                  \end{aligned}
                              \end{align*}
                    \end{itemize}

              \item for SP-image:
                    \begin{align*}
                        \begin{aligned}
                             & t_{single} = 554.0763 - 2.9193 n_{batch}            \\
                             & \mbox{ with adjusted } R^2 = -0.2175, SE = 166.8644
                        \end{aligned}
                    \end{align*}
          \end{itemize}
\end{itemize}

\paragraph{Comments}

For JS-block and JS-grouping, $ y = a + b/x + c/x^2 $ is the best-fitting model.
For SP-image, all the models perform poorly because of the noisy and sparse samples.

For $ y = a $, JS-block data fits okay with $ 1 \times 1 $ and $ 8 \times 8 $ being two significant outliers.
As explained above, for $ 1 \times 1 $, $ t_{single} $ approaches 0 because single edit and view are more concurrent than in other layouts.
For $ 8 \times 8 $, the large $ t_{single} $ is likely due to the latency of the interface.
JS-grouping data fits well, with $ 1 \times 1 $ being the major outlier, and the reason is the same as for JS-block.
SP-image data fits poorly.
Two major outliers are $ 2 \times 3 $ and $ 3 \times 4 $.
However, note that the deviations at these two points are roughly the negation of each other.
Thus, it is possible that $ y = a $ suits the data.

For $ y = a + bx $, for all three applications, the slopes are not too steep, and thus the analysis is similar to that of $ y = a $.

\paragraph{Outcome}

We finally decide to use the simple constant function $ y = a $ because:

\begin{itemize}
    \item It makes sense logically.
          When there are more grid cells, the distance between grid cells decreases.
          The reduced distance makes it easier to conduct single edit.
          Meanwhile, the size of grid cells gets smaller, which makes it hard to single edit.
          These two effects may offset.

    \item Although $ y = a + b/x + c/x^2 $ numerically fits JS-block and JS-grouping well, it fits SP-image poorly and thus may not be reliable.

    \item It is the simplest model and may avoid overfitting.
\end{itemize}

Besides, we decide not to discard the outlying points, as they do not change the resulting model much.
The functions we choose are:

\begin{itemize}
    \item \textbf{for JS-block:} $\mathbf{t_{single} = 208.0818}$
          \[ \mbox{ with adjusted } R^2 = 0, SE = 95.4480 \]

    \item \textbf{for JS-grouping:} $\mathbf{t_{single} = 193.8312}$
          \[ \mbox{ with adjusted } R^2 = 0, SE = 70.3764 \]

    \item \textbf{for SP-image:} $\mathbf{t_{single} = 507.9517}$
          \[ \mbox{ with adjusted } R^2 = 0, SE = 151.2289 \]
\end{itemize}

\subsection{Reestimate \texorpdfstring{$\mathbf{t_{view}}$}{t\_view} by \texorpdfstring{$\mathbf{t_{new}}$}{t\_new} and \texorpdfstring{$\mathbf{t_{single}}$}{t\_single} Models}

\paragraph{Goal}

Reestimate $ t_{view} $ by removing the previous modeled $ t_{new} $ and $ t_{single} $ from the equations to possibly reduce the noise and make sure the original equations still approximately hold.

\paragraph{Procedure}

With the $t_{new}$ and $t_{single}$ modeled in the last sections, we reestimate $t_{view}$ from the experiment data by removing the contribution of $t_{new}$ and $t_{single}$ to equations as

\begin{gather}
    \begin{bmatrix}
        t_{view}
    \end{bmatrix}
    = (X^T X)^{-1} X^T
    \begin{bmatrix}
        T_1 - N_{new, 1}t_{new} - N_{single, 1}t_{single} \\ T_2 - N_{new, 2}t_{new} - N_{single, 2}t_{single} \\ ... \\ T_{n} - N_{new, n}t_{new} - N_{single, n}t_{single}
    \end{bmatrix}
\end{gather}
where X is the $N_{view}$ measured for the $ n $ trials
\begin{align*}
    X =
    \begin{bmatrix}
        N_{view, 1} \\
        N_{view, 2} \\
        ...         \\
        N_{view, n} \\
    \end{bmatrix}
\end{align*}
and the $t_{new}$ and $t_{single}$ in the formula use modeled $t_{new}$ and $t_{single}$ values in the last sections.

\paragraph{Outcome}

\begin{figure}[htbp]
    \centering
    \includegraphics[width=\linewidth]{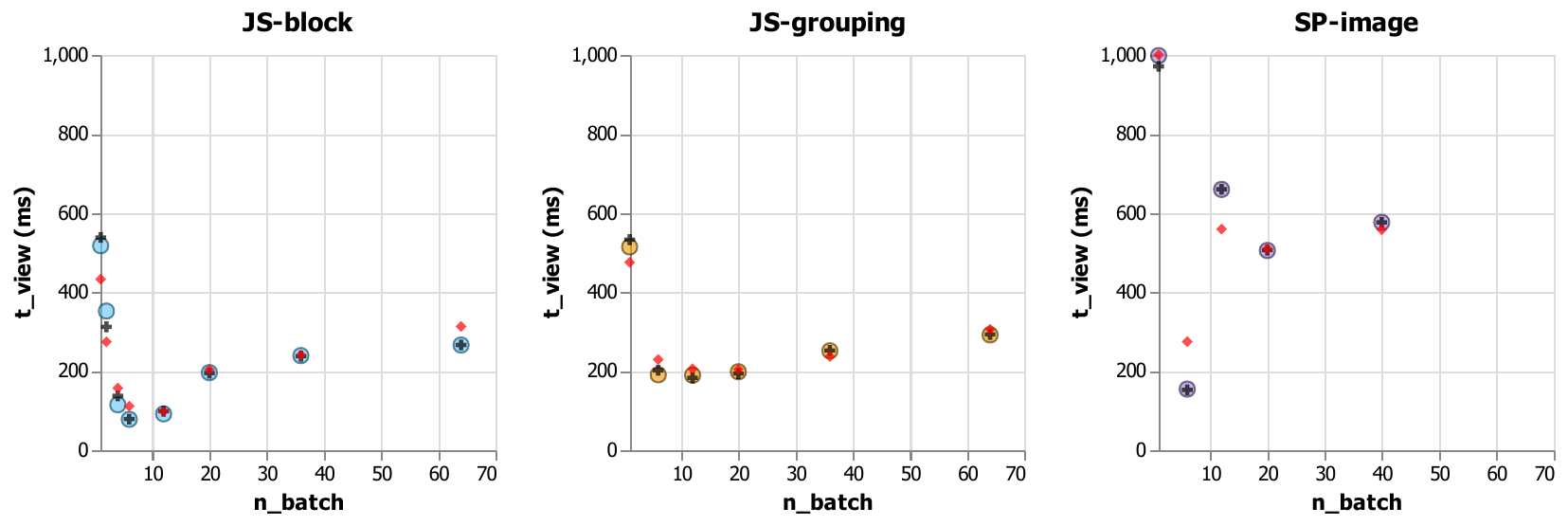}
    \caption{
        $t_{view}$ reestimation by $t_{new}$ and $t_{single}$:
        The influence of reestimation to $t_{view}$ as a function of $n_{batch}$ for the three applications.
        The colored dots are initial estimations by multiple linear regression.
        The black crosses are reestimations by putting only $ t_{new} $ back.
        The red diamonds are reestimations by putting both $ t_{new} $ and $ t_{single} $ back.
    }
    \Description[
        Fully described in the caption.
    ]{}
    \label{fig:t-view-reestimate-2}
\end{figure}

Figure~\ref{fig:t-view-reestimate-2} shows the influence of the reestimation on $t_{view}$.

\paragraph{Comments}

The modeling of $t_{new}$ does not significantly change the estimation of $t_{view}$.
By comparison, the additional modeling of $t_{single}$ changes $t_{view}$ observably.

For JS-block, the $ 1 \times 1 $ point lowers significantly after reestimation for $ t_{new} $ and $ t_{single} $ because the modeled $ t_{single} $ at $ 1 \times 1 $ is much larger than the initial $ t_{single} $ which is close to 0.
The reestimation decreases the $ t_{view} $ at $ 1 \times 1 $ to compensate for this change.

Similarly, for JS-grouping, the $ 1 \times 1 $ point lowers significantly.
For SP-image, the $ 2 \times 3 $ point increases significantly while the $ 3 \times 4 $ point decreases significantly.

All the other data points are hardly changed.

\subsection{Model Reestimated \texorpdfstring{$\mathbf{t_{view}}$}{t\_view}}

\paragraph{Goal}

Model the reestimated $ t_{view} $ to reduce noise.

\paragraph{Procedure}

We model $t_{view}$ as a function of $n_{batch}$.
We fit a model for each application because we expect viewing time to be related to the content.
For JS-block, there are 8 data points.
For JS-grouping, there are 6 data points.
For SP-image, there are 5 data points.
We assume the suitable model function family should be shared among applications.
We have used the Nonlinear Regression Tool~\cite{Xuru2006Online} to produce a set of candidate model functions $ t_{new} = f(n_{batch}) $.

\begin{itemize}
    \item With the number of parameters = 2, the top 3 functions are

          \begin{itemize}
              \item for JS-block:
                    \begin{itemize}
                        \item $y = ax^{b/x} \mbox{ with } R^2 = 0.5548$
                        \item $y = \frac{ax}{x + b} \mbox{ with } R^2 = 0.5397$
                        \item $y = ae^{b/x} \mbox{ with } R^2 = 0.5098$
                    \end{itemize}

              \item for JS-grouping:
                    \begin{itemize}
                        \item $y = \frac{ax}{x + b} \mbox{ with } R^2 = 0.8494$
                        \item $y = ae^{b/x} \mbox{ with } R^2 = 0.8477$
                        \item $y = a + b/x \mbox{ with } R^2 = 0.8174$
                    \end{itemize}

              \item for SP-image:
                    \begin{itemize}
                        \item $y = ax^{b/x} \mbox{ with } R^2 = 0.8815$
                        \item $y = \frac{ax}{x + b} \mbox{ with } R^2 = 0.7500$
                        \item $y = ae^{b/x} \mbox{ with } R^2 = 0.7289$
                    \end{itemize}
          \end{itemize}

    \item With the number of parameters = 3, the top 3 functions are

          \begin{itemize}
              \item for JS-block:
                    \begin{itemize}
                        \item $y = a + bx + c/x \mbox{ with } R^2 = 0.9310$
                        \item $y = x^ae^{bx^c} \mbox{ with } R^2 = 0.9093$
                        \item $y = a + bx + c/x^2 \mbox{ with } R^2 = 0.8822$
                    \end{itemize}

              \item for JS-grouping:
                    \begin{itemize}
                        \item $y = a + bx + c/x \mbox{ with } R^2 = 0.9958$
                        \item $y = x^ae^{bx^c} \mbox{ with } R^2 = 0.9886$
                        \item $y = ax^be^{cx} \mbox{ with } R^2 = 0.9874$
                    \end{itemize}

              \item for SP-image:
                    \begin{itemize}
                        \item $y = a + \frac{b}{x + c} \mbox{ with } R^2 = 0.9803$
                        \item $y = ae^{\frac{b}{x + c}} \mbox{ with } R^2 = 0.9795$
                        \item $y = a + b/x + c/x^2 \mbox{ with } R^2 = 0.9546$
                    \end{itemize}
          \end{itemize}

    \item With the number of parameters = 4, the top 3 functions are

          \begin{itemize}
              \item for JS-block:
                    \begin{itemize}
                        \item $y = a + bx + c/x + dln(x) \mbox{ with } R^2 = 0.9378$
                        \item $y = a + bx + c/x + d/x^2 \mbox{ with } R^2 = 0.9311$
                        \item $y = a + bx + c/x \mbox{ with } R^2 = 0.9310$
                    \end{itemize}

              \item for JS-grouping:
                    \begin{itemize}
                        \item $y = a + bx + c/x + dln(x) \mbox{ with } R^2 = 0.9998$
                        \item $y = a + bx + c/x + d/x^2 \mbox{ with } R^2 = 0.9997$
                        \item $y = ae^{b/x} + ce^{d/x} \mbox{ with } R^2 = 0.9993$
                    \end{itemize}

              \item for SP-image:
                    \begin{itemize}
                        \item $y = \frac{a + bx + cx^2}{x + d} \mbox{ with } R^2 = 0.9862$
                        \item $y = a + \frac{b}{x + c} \mbox{ with } R^2 = 0.9803$
                        \item $y = ae^{\frac{b}{x + c}} \mbox{ with } R^2 = 0.9795$
                    \end{itemize}
          \end{itemize}
\end{itemize}

It can be seen that $ y = a + bx + c/x $ is reoccurring.
It is the best for JS-block and JS-grouping when \#parameters = 3 and the third-best for JS-block when \#parameters = 4.
Adding $ dln(x) $ and $ d/x^2 $ terms generate the best and second best functions for JS-block and JS-grouping when \#parameters = 4.
Removing the $ cx $ term generates the third-best for JS-grouping when \#parameters = 2.
For SP-image, the data is sparse and noisy and thus not decisive.

Thus, we choose $ y = a + bx + c/x $ as the function family for $ t_{view} $.
We fit the model with linear regression.

\paragraph{Outcome}

\begin{figure}[htbp]
    \centering
    \includegraphics[width=\linewidth]{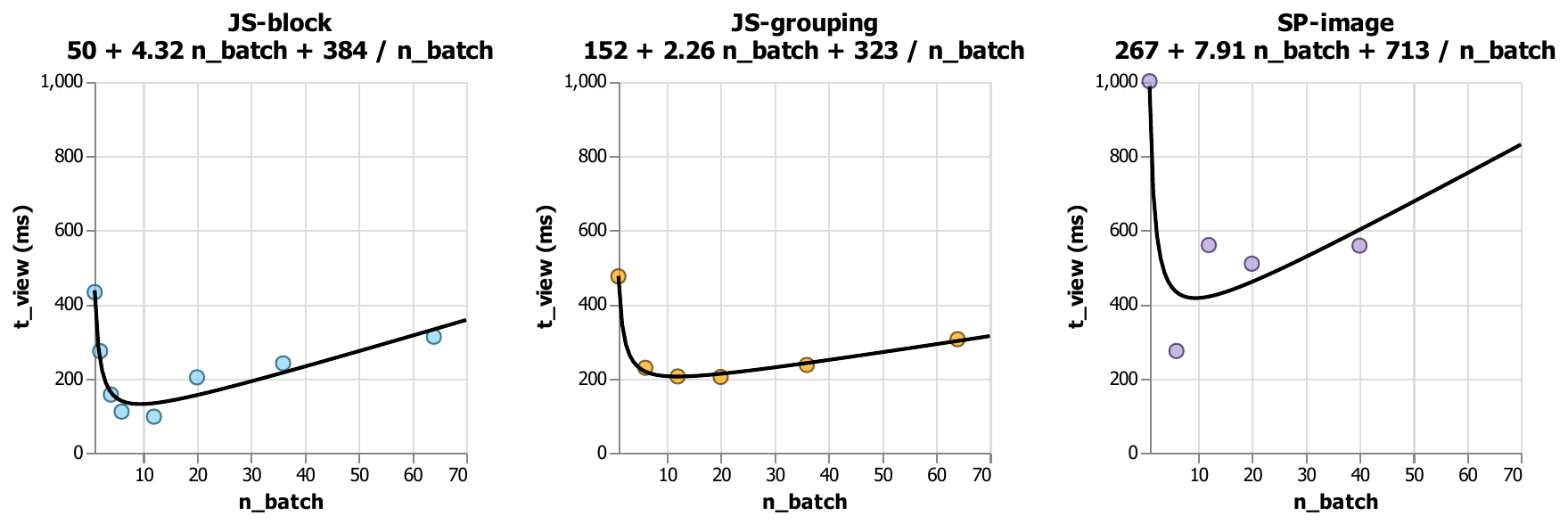}
    \caption{
        The fitted $t_{view}$ as a function of $n_{batch}$ for the three applications using $y=a + bx + c/x$.
    }
    \Description[
        Fully described in the main text of the section.
    ]{}
    \label{fig:model-t-view}
\end{figure}

Figure~\ref{fig:model-t-view} shows the fitting result for $ t_{view} $.
The functions fitted are:

\begin{itemize}
    \item \textbf{for JS-block:} $\mathbf{t_{view} = 50.4283 + 4.3202 n_{batch} + \frac{383.5033}{n_{batch}}}$
          \[ \mbox{ with adjusted } R^2 = 0.9034, SE = 34.8005 \]

    \item \textbf{for JS-grouping:} $\mathbf{t_{view} = 151.9951 + 2.2618 n_{batch} + \frac{322.6178}{n_{batch}}}$
          \[ \mbox{ with adjusted } R^2 = 0.9930, SE = 8.7040 \]

    \item \textbf{for SP-image:} $\mathbf{t_{view} = 267.0109 + 7.9103 n_{batch} + \frac{712.8504}{n_{batch}}}$
          \[ \mbox{ with adjusted } R^2 = 0.6472, SE = 156.1362 \]
\end{itemize}

\paragraph{Comments}

The data clearly shows a U-shaped pattern.
We conjecture that small grid cells require more attention and time to see clearly, while large grid cells can be distracting and require more eye movement to view one object.

$n_{batch} \cdot gridCellArea  = interfaceArea $ is a constant.
If we assume the penalty of time cost to look into details of small objects to be inversely proportional to object area, the $bx$ term can be explained.
Assuming the eye moment time to scan an object is proportional to the object's area, the $c/x$ term can be explained.
The viewing action may contain some reaction latency of the human.
Thus, the $a$ term can be explained.

\subsection{Final Estimations of Operator Time Costs}

Table~\ref{tab:experiment-summary-reestimate} shows the final estimations of operator time costs with the models we have fitted for $t_{new}$, $t_{view}$, and $t_{single}$.

\begin{table}[htbp]
    \centering
    \footnotesize
    \caption{
        Modeled operator time costs: The final estimations of operator time costs.
    }
    \begin{tabular}{ccccc}
        \toprule
        \textbf{Application} & \textbf{Layout} & $\mathbf{t_{new}}$ (ms) & $\mathbf{t_{view}}$ (ms) & $\mathbf{t_{single}}$ (ms) \\
        \midrule
        JS-block             & 1 $\times$ 1    & 443                     & 438                      & 208                        \\
        \midrule
        JS-block             & 1 $\times$ 2    & 515                     & 251                      & 208                        \\
        \midrule
        JS-block             & 2 $\times$ 2    & 691                     & 164                      & 208                        \\
        \midrule
        JS-block             & 2 $\times$ 3    & 769                     & 140                      & 208                        \\
        \midrule
        JS-block             & 3 $\times$ 4    & 859                     & 134                      & 208                        \\
        \midrule
        JS-block             & 4 $\times$ 5    & 897                     & 156                      & 208                        \\
        \midrule
        JS-block             & 6 $\times$ 6    & 924                     & 217                      & 208                        \\
        \midrule
        JS-block             & 8 $\times$ 8    & 939                     & 333                      & 208                        \\
        \midrule
        JS-grouping          & 1 $\times$ 1    & 443                     & 477                      & 194                        \\
        \midrule
        JS-grouping          & 2 $\times$ 3    & 769                     & 219                      & 194                        \\
        \midrule
        JS-grouping          & 3 $\times$ 4    & 859                     & 206                      & 194                        \\
        \midrule
        JS-grouping          & 4 $\times$ 5    & 897                     & 213                      & 194                        \\
        \midrule
        JS-grouping          & 6 $\times$ 6    & 924                     & 242                      & 194                        \\
        \midrule
        JS-grouping          & 8 $\times$ 8    & 939                     & 302                      & 194                        \\
        \midrule
        SP-image             & 1 $\times$ 1    & 443                     & 988                      & 508                        \\
        \midrule
        SP-image             & 2 $\times$ 3    & 769                     & 433                      & 508                        \\
        \midrule
        SP-image             & 3 $\times$ 4    & 859                     & 421                      & 508                        \\
        \midrule
        SP-image             & 4 $\times$ 5    & 897                     & 461                      & 508                        \\
        \midrule
        SP-image             & 5 $\times$ 8    & 927                     & 601                      & 508                        \\
        \bottomrule
    \end{tabular}%
    \label{tab:experiment-summary-reestimate}%
\end{table}%

  \ifx\hidemain\undefined
  \else
    \bibliographystyle{style/ACM-Reference-Format}
    \bibliography{assets/bibs/papers.bib}
  \fi
\fi

\end{document}